\documentclass[final,1p,times]{elsarticle}

\usepackage{amssymb}

\usepackage{enumitem}

\usepackage{caption}
\usepackage{subcaption}

\usepackage{amsmath,amsthm,mathtools}
\usepackage[hyphens]{url}

\begin{document}

\begin{frontmatter}

\title{Calibrations of the Compton Spectrometer and Imager}

\author[1]{Jacqueline Beechert\corref{cor1}}
\ead{jbeechert@berkeley.edu}

\author[1]{Hadar Lazar}

\author[2]{Steven E. Boggs}

\author[3]{Terri J. Brandt}

\author[4]{Yi-Chi Chang}

\author[4]{Che-Yen Chu}

\author[1]{Hannah Gulick}

\author[3]{Carolyn Kierans}

\author[1]{Alexander Lowell}

\author[1]{Nicholas Pellegrini}

\author[2]{Jarred M. Roberts}

\author[5,6,2]{Thomas Siegert}

\author[7]{Clio Sleator}

\author[1]{John A. Tomsick}

\author[1]{Andreas Zoglauer}

\address[1]{Space Sciences Laboratory, University of California, Berkeley, 7 Gauss Way, Berkeley, CA 94720, USA}

\address[2]{Center for Astrophysics and Space Sciences, University of California, San Diego, 9500 Gilman Dr., La Jolla, CA 92093, USA}

\address[3]{NASA Goddard Space Flight Center, Greenbelt, MD 20771, USA}

\address[4]{National Tsing Hua University, 101, Section 2, Kuang-Fu Road, Hsinchu 300044, Taiwan R.O.C.}

\address[5]{Max-Planck-Institute for extraterrestrial Physics, Giessenbachstr. 1, 85748, Garching bei M\"{u}nchen, Germany}

\address[6]{Institut f\"{u}r Theoretische Physik und Astrophysik, Universit\"{a}t W\"{u}rzburg, Campus Hubland Nord, Emil-Fischer-Str. 31, 97074 W\"{u}rzburg, Germany}

\address[7]{U.S. Naval Research Laboratory, Washington DC 20375, USA}

\cortext[cor1]{Corresponding author}

\journal{NIM A}

\begin{abstract}
The Compton Spectrometer and Imager (COSI) is a balloon-borne soft $\gamma$-ray telescope (0.2--5\,MeV) designed to study astrophysical sources. COSI employs a compact Compton telescope design and is comprised of twelve high-purity germanium semiconductor detectors. Tracking the locations and energies of $\gamma$-ray scatters within the detectors permits high-resolution spectroscopy, direct imaging over a wide field-of-view, polarization studies, and effective suppression of background events. Critical to the precise determination of each interaction's energy, position, and the subsequent event reconstruction are several calibrations conducted in the field before launch. Additionally, benchmarking the instrument's higher-level performance through studies of its angular resolution, effective area, and polarization sensitivity quantifies COSI's scientific capabilities. In May 2016, COSI became the first science payload to be launched on NASA's superpressure balloon and was slated for launch again in April 2020. Though the 2020 launch was canceled due to the COVID-19 pandemic, the COSI team took calibration measurements prior to cancellation. In this paper we provide a detailed overview of COSI instrumentation, describe the calibration methods, and compare the calibration and benchmarking results of the 2016 and 2020 balloon campaigns. These procedures will be integral to the calibration and benchmarking of the NASA Small Explorer satellite version of COSI scheduled to launch in 2025. 
\end{abstract}


\end{frontmatter}


\section{Introduction}
\label{sec:introduction}

\begin{figure}
\centering
\begin{minipage}{0.48\textwidth}
\centering
\includegraphics[height=5cm]{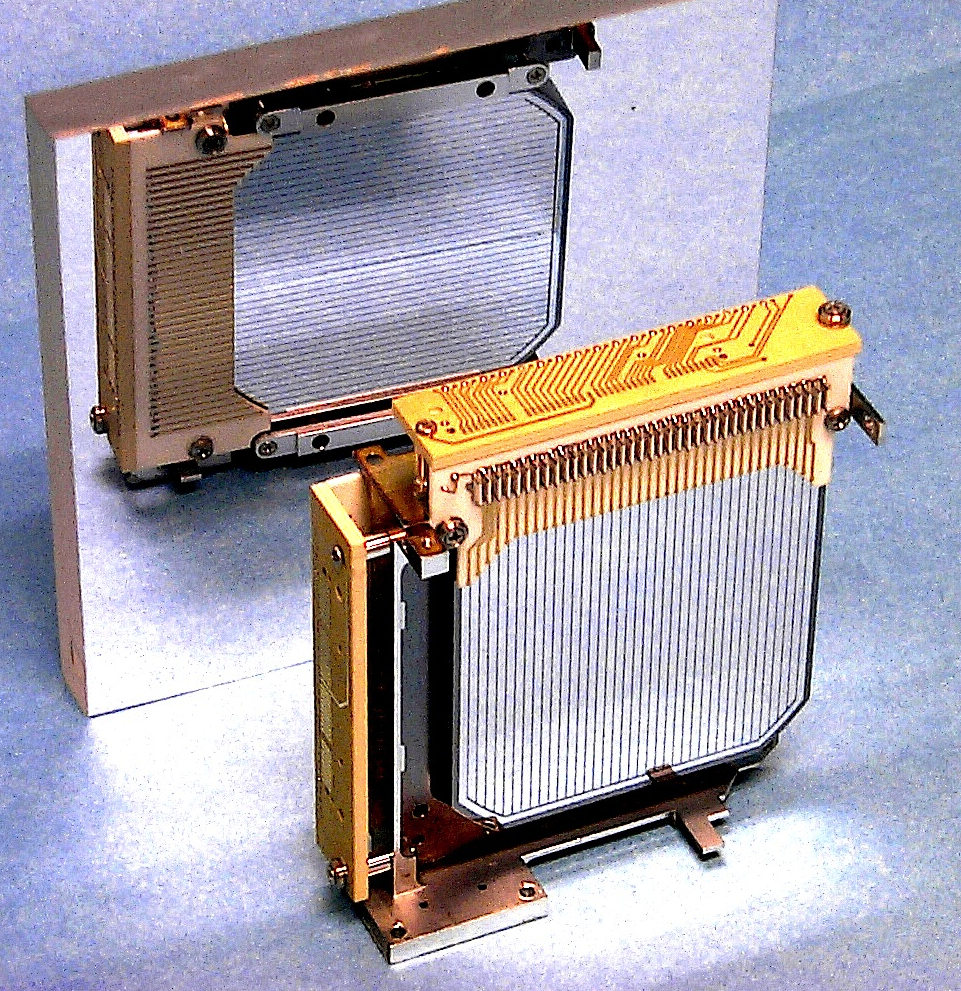}
\caption{One of COSI's twelve double-sided strip GeDs. Reflection in mirror shows orthogonal strips on opposite faces of the detector. Image from \cite{SLEATOR2019162643}.}
\label{fig:gedmirror}
\end{minipage}
\hfill
\begin{minipage}{0.48\textwidth}
\includegraphics[width =0.95\textwidth]{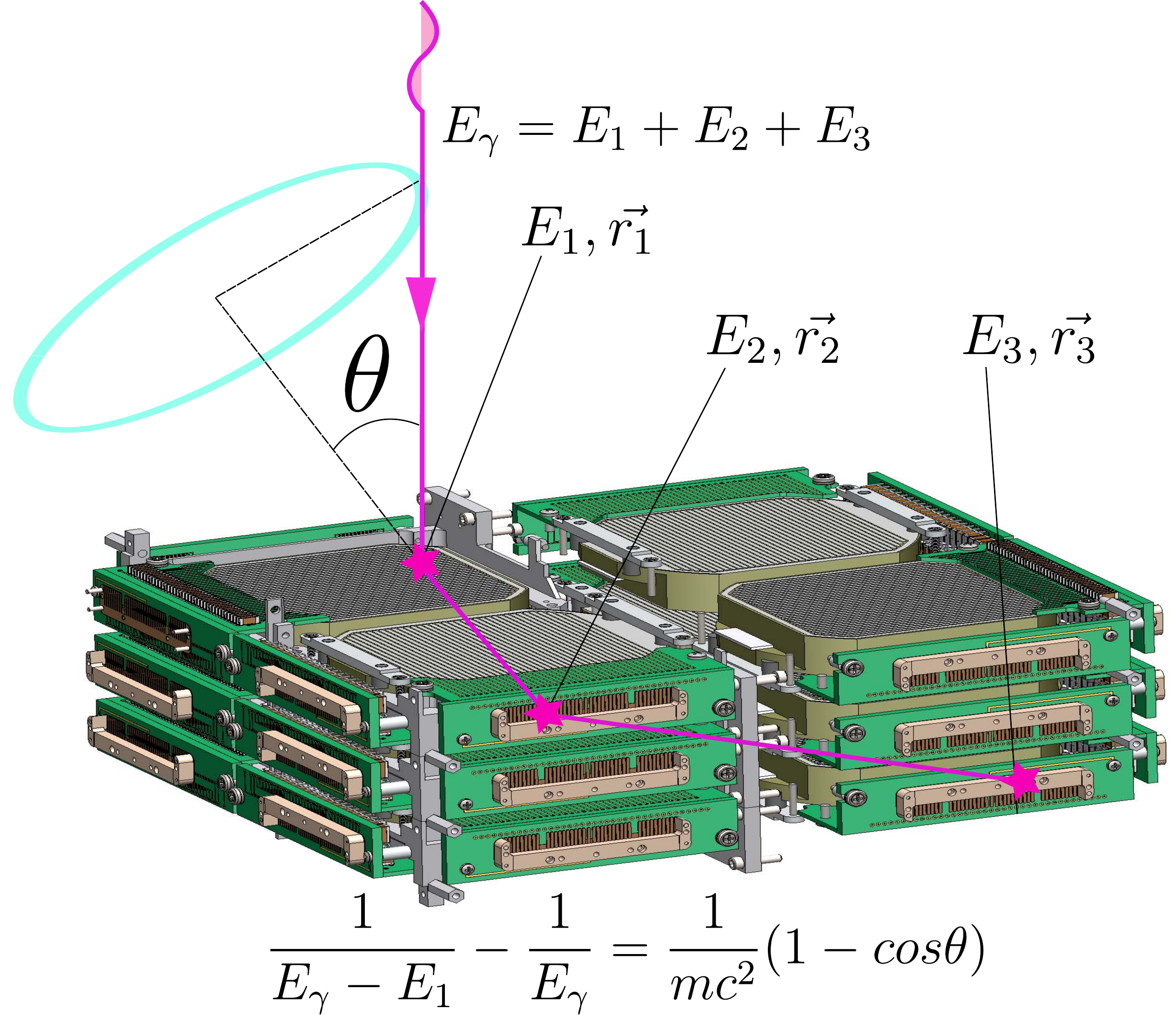}
\caption{Schematic of the COSI detector stack and the Compton principle. The labels $\vec{r_i}$, $E_i$ indicate the position and energy of the $i$th Compton scatter of an incident photon with energy $E_{\gamma}$. Image from \cite{lowell:2016}.}
\label{fig:comptonprinciple}
\end{minipage}
\end{figure}

The Compton Spectrometer and Imager (COSI) is a balloon-borne compact Compton telescope (CCT) uniquely equipped to probe the ``MeV gap" (0.1--10\,MeV) \cite{2019BAAS...51g.245M} of $\gamma$-ray astrophysics. Though historically under-explored due to experimental challenges inherent to its energy range and low interaction cross-sections, the MeV gap is rich in scientific potential. COSI's main science goals in this bandpass are to determine the nature of Galactic positrons, measure emission from Galactic nucleosynthesis, and perform novel polarization measurements of gamma-ray bursts (GRBs) and compact objects. Its twelve high-purity germanium detectors (GeDs; Figure\,\ref{fig:gedmirror}) facilitate studies of the 0.2--5\,MeV energy range through reconstruction of incident $\gamma$-rays' Compton interactions inside the active detector volume.
Specifically, Compton reconstruction uses interaction locations and energies to constrain the origins of incident photons to an annulus on the sky (Figure\,\ref{fig:comptonprinciple}). Compton reconstruction is inherently sensitive to photon polarization and is an effective method of background reduction  \cite{2000A&AS..145..311B,zoglauer2021cosi}.

COSI's excellent spectral resolution, wide-field direct imaging capabilities, and ability to perform Compton polarimetry distinguish it as a powerful observatory for soft $\gamma$-ray science. As a successor to the Nuclear Compton Telescope (NCT) \cite{Bandstra_2011}, COSI draws upon demonstrated scientific achievement in CCTs. COSI was redesigned from NCT for compatibility with NASA's superpressure balloon (SPB) and in 2016 became the first science payload to fly on the latest SPB technology \cite{2016int..workE..75K}. Launched from Wanaka, New Zealand on May 17, 2016, COSI remained afloat for 46 days before terminating safely over Peru on July 2, 2016. A complete overview of the flight is provided in \cite{2016int..workE..75K}. COSI observed the 511\,keV annihilation signature at the Galactic Center \cite{kierans2020detection,siegert2020imaging}, set a constraining upper limit on the polarization of GRB160530A \cite{lowell2017polarimetric}, detected the Crab pulsar \cite{sleator2019measuring}, and measured Galactic $\mathrm{^{26}Al}$ \cite{Beechert:2021tZ,beechert2022measurement}. Studies of other MeV point sources, such as Centaurus A, Cygnus X-1, and the Vela pulsar, are currently underway.

In order to understand the many intricacies of the Compton measurement process for these science goals, it is necessary to characterize the instrument via calibrations and response checks. The calibration procedures and studies of the instrument response detailed in this paper are directly relevant to future missions employing double-sided cross-strip GeDs, which likewise require energy and positional reconstruction as well as an understanding of detector-wide effects. These missions include the next generation of COSI as a NASA Small Explorer satellite\footnote{NASA press release announcing the selection of COSI as a satellite mission: https://www.nasa.gov/press-release/nasa-selects-gamma-ray-telescope-to-chart-milky-way-evolution}, which carries significant design heritage from the COSI balloon mission, and the Gamma-ray Imager/Polarimeter for Solar Flares (GRIPS) balloon instrument \cite{2016SPIE.9905E..2QD}. The former is slated for launch in 2025 and the latter in 2024.

This paper is structured as follows: in Section\,\ref{sec:instrument} we introduce the COSI instrument. We provide an overview of the MEGAlib analysis pipeline used to calibrate and benchmark COSI's performance in Section\,\ref{sec:megalib}, and in Section\,\ref{sec:calibrations} we detail the calibration procedures and results from the 2016 and 2020 calibration campaigns. In Section\,\ref{sec:instrument_performance} we analyze instrument performance and we conclude in Section\,\ref{sec:conclusion}. The instrument design and calibration procedures in this paper were developed primarily at the Space Sciences Laboratory (SSL) in Berkeley, CA.

\section{COSI Instrument}
\label{sec:instrument}

\subsection{3-D Position Sensitive Germanium Detectors}

The heart of COSI is comprised of twelve double-sided cross-strip high-purity GeDs. Each measuring $8 \times 8 \times 1.5\,\mathrm{cm^3}$, the GeDs were developed using the Lawrence Berkeley National Laboratory's amorphous Ge contact technology \cite{amman2020high}. Each side of the detectors is instrumented with 37 aluminum strip electrodes of 2\,mm strip pitch deposited orthogonally on the anode and cathode (Figure\,\ref{fig:gedmirror}). A gap between strips of 0.25\,mm was chosen to strike a balance between small gaps, which minimize the loss of charge carriers that fall between strips, and large gaps, which improve energy resolution through decreased strip capacitance. A 2\,mm wide guard ring surrounds the active area of each detector face to reject events close to the edge of the detector and to minimize leakage current. 

The 888 total strips define COSI's three-dimensional position sensitivity: the $x$--$y$ position of a photon interaction is determined by the intersection of orthogonal triggered strips \cite{phlips04} and the $z$-position is determined through the timing difference between the collection of electrons on the anode and holes on the cathode (see Section\,\ref{sec:depth_calibration} for a detailed discussion of this timing-to-depth calibration). The 3-D position resolution, defined as the product of the $x-$, $y-$, and $z-$position resolutions ($\sim$2\,mm, $\sim$2\,mm, and $\sim$0.5\,mm, respectively), is approximately 2\,mm$^{3}$ \cite{bandstra_thesis}.

\begin{figure}
\centering
\begin{minipage}{0.48\textwidth}
\centering
\includegraphics[width = 0.9\textwidth]{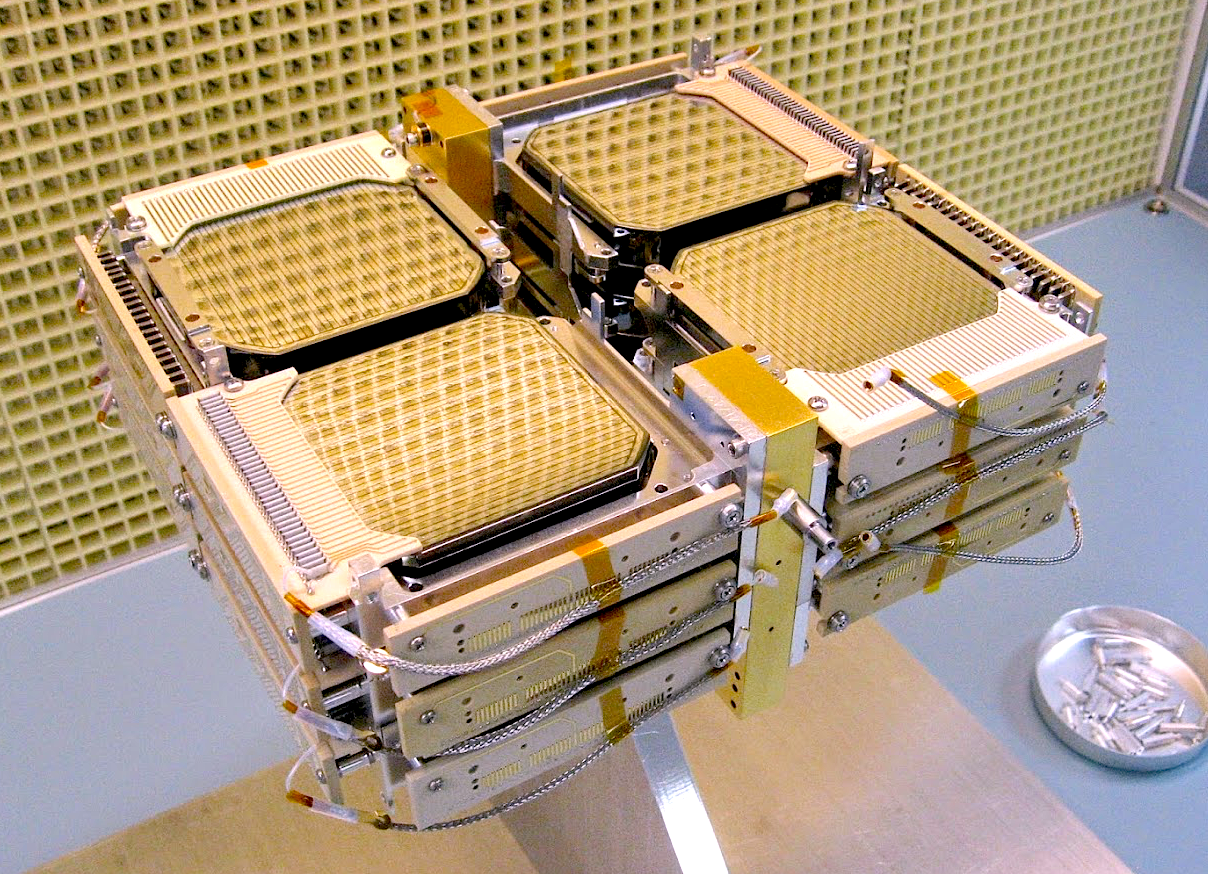}
\caption{Photograph of the COSI detector stack before integration into the cryostat. Image from \cite{2016int..workE..75K}.}
\label{fig:detectorstack}
\end{minipage}
\hfill
\begin{minipage}{0.48\textwidth}
\centering
\includegraphics[width = 0.87\textwidth]{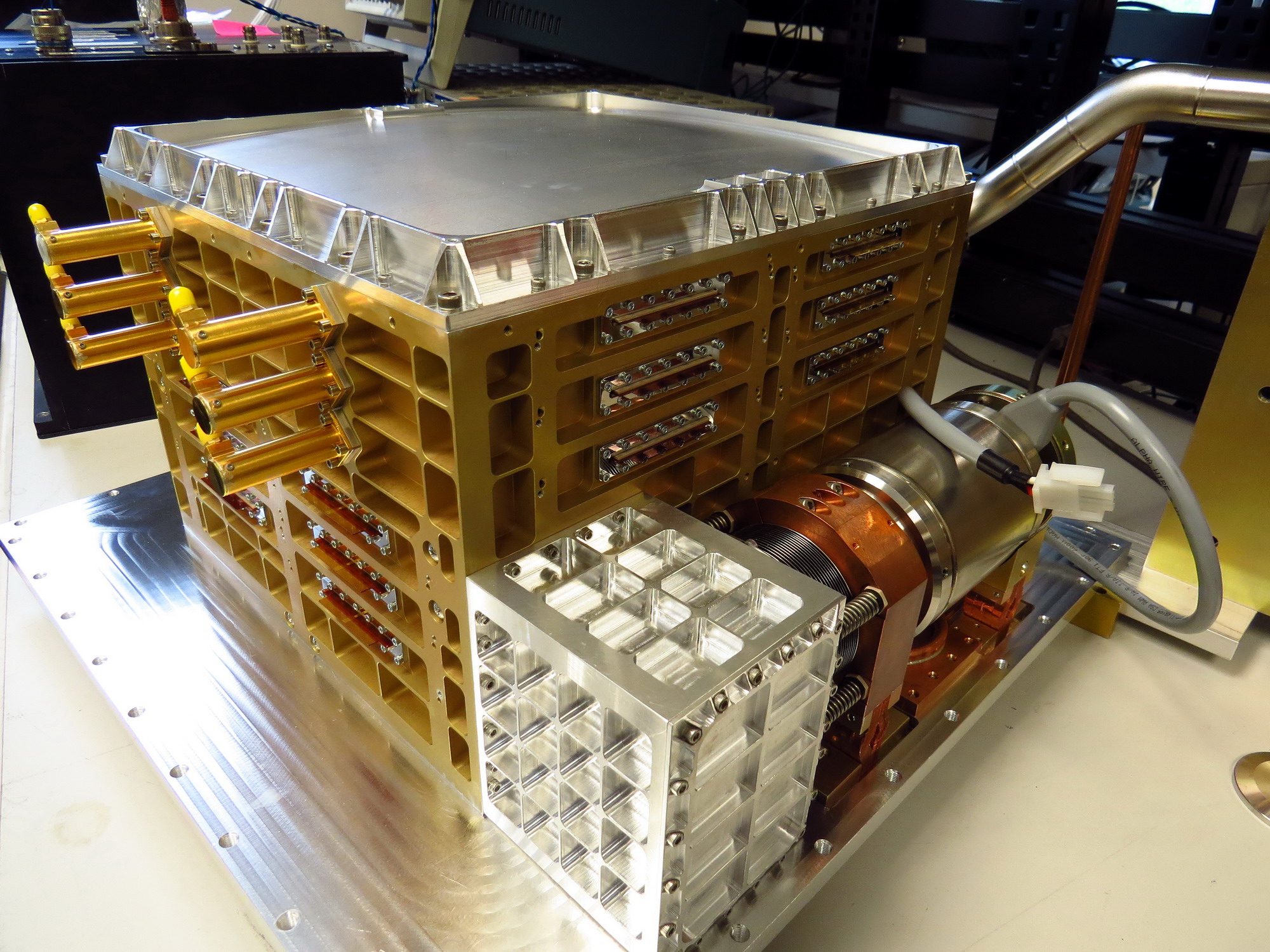}
\caption{The closed COSI cryostat houses the detector stack. The cryocooler is visible in the foreground.}
\label{fig:prettycryostat}
\end{minipage}
\end{figure}

The detectors are stacked in a $2 \times 2 \times 3$ configuration (Figure\,\ref{fig:detectorstack}) and are over-depleted with bias voltages between 1000 and 1500 V. We apply AC coupling to the high-voltage side and consequently refer to the high-voltage side of each detector as the ``AC side" (anode) and the low-voltage side as the ``DC side" (cathode). To mitigate the effects of electronic noise, coincidence triggers on both the AC and DC sides of a detector are required to trigger the readout (see Section\,\ref{sec:readout} for more details).

\subsection{Detector Head}
\label{sec:detector_head}

The twelve GeDs are housed in an anodized aluminum cryostat evacuated to pressures of approximately 10$^{-6}$ Torr (Figure\,\ref{fig:prettycryostat}). COSI's GeDs operate at cryogenic temperatures because thermal excitations from warmer temperatures would exceed germanium's small bandgap ($\sim$0.7\,eV at 300\,K) and cause prohibitive leakage current. To this end, the GeDs are cooled with a Sunpower CryoTel CT mechanical cryocooler. Mechanically cooling the instrument is a preferred alternative to cooling with consumable liquid nitrogen because storing liquid nitrogen would add considerable weight to the instrument and limit the operational lifetime. The cryocooler is compactly located outside of the cryostat (Figure\,\ref{fig:prettycryostat}) and operates continuously throughout flight.

During the COSI 2016 flight, the cryocooler was run in ``constant temperature" mode, which set the target cryocooler coldtip temperature to 77\,K and expended about 100\,W. During COSI 2020 ground calibrations, ``constant power" mode kept the cryocooler at a constant power of 95\,W in an attempt to stabilize power-dependent cryocooler vibrations. In practice, this change did not impact instrument performance and we find consistency between 2016 and 2020 calibrations. In both modes, thermal losses in the coupling of the cryocooler cold tip to each of the twelve GeDs resulted in detector temperatures of $\sim$ 83--84\,K. The temperature of the cryocooler itself is regulated with an external high-power computer fan on the ground and with an active liquid cooling system during flight to avoid overheating \cite{sleator2019measuring}. 

\begin{figure}
\centering
\includegraphics[width = 0.6\textwidth]{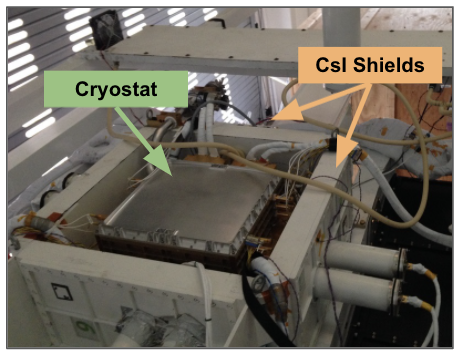}
\caption{COSI detector head: the CsI shields surround the cryostat.}
\label{fig:detectorhead}
\end{figure}

The cryostat is surrounded on four sides and the bottom by six scintillator detectors (shields), each comprised of a $40 \times 20 \times 4$\,cm$^3$ block of cesium iodide (CsI) (Figure\,\ref{fig:detectorhead}). The signals from each shield are read out by two photomultiplier tubes (PMTs) with an energy threshold of $\sim$ 80\,keV. The signals from the twelve total PMTs are OR'ed together as one veto pulse that, if high 0.7--1.1\,$\mu$s after a GeD signal, vetoes the event from the analysis \cite{sleator2019measuring}. Thus, the CsI anticoincidence shields reduce background by limiting the field of view to 25\% of the sky ($\sim$ $\pi$ sr) at one time and rejecting incompletely absorbed events which cannot be reconstructed.

The COSI cryostat and CsI shields are mounted on top of a $5 \times 5 \times 7$\,feet$^{3}$ lightweight, three-tier, aluminum gondola frame. For more details about control of the gondola during flight, including the flight computer, power systems, telemetry, and GPS, refer to \cite{lowell:phdthesis}.

\subsection{Readout Electronics}
\label{sec:readout}

COSI uses low-power, low-noise, custom analog electronics for signal processing. Each of the 888 detector strips has its own readout chain. First, the electrode signals are fed through the cryostat walls on Kapton-manganin flex circuits. A total of 24 low-power, low-noise, charge sensitive preamplifier boxes mounted to the sides of the cryostat perform the signal extraction. One preamplifier box reads the signal from one detector side and consumes 0.5\,W of power. The signal from the preamplifiers is subsequently transferred with teflon-insulated, 50\,$\Omega$ controlled impedance coaxial ribbon cables made by Gore to a pulse-shaping amplifier with fast and slow channels. The fast channel uses a small delay line constant fraction discriminator to time stamp each waveform at 50\% of its maximum amplitude, generating a low time walk signal. The slow channel, with a $\sim$20\,keV threshold, uses a unipolar shaper with a 6\,$\mu$s shaping time for noise reduction and accurate pulse height determination.

Twelve ``card cages" house the pulse processing and triggering electronics for each detector. Each card cage contains eight analog boards with pulse-shaping circuits and digital logic, a high-voltage power supply board to bias the GeDs with 1000--1500 V, a low-voltage power supply board to power the rail voltages in the card cage electronics, and an FPGA board that retrieves pulse height and timing information from the analog boards. All boards are connected to the same backplane which supplies bi-directional housekeeping communication, power, and event data channels. The total power consumed by a card cage is $\sim$ 20\,W.

\section{MEGAlib Software Pipeline}
\label{sec:megalib}

\begin{figure}
\centering
\begin{minipage}{0.45\textwidth}
\centering
\includegraphics[width = 1\textwidth]{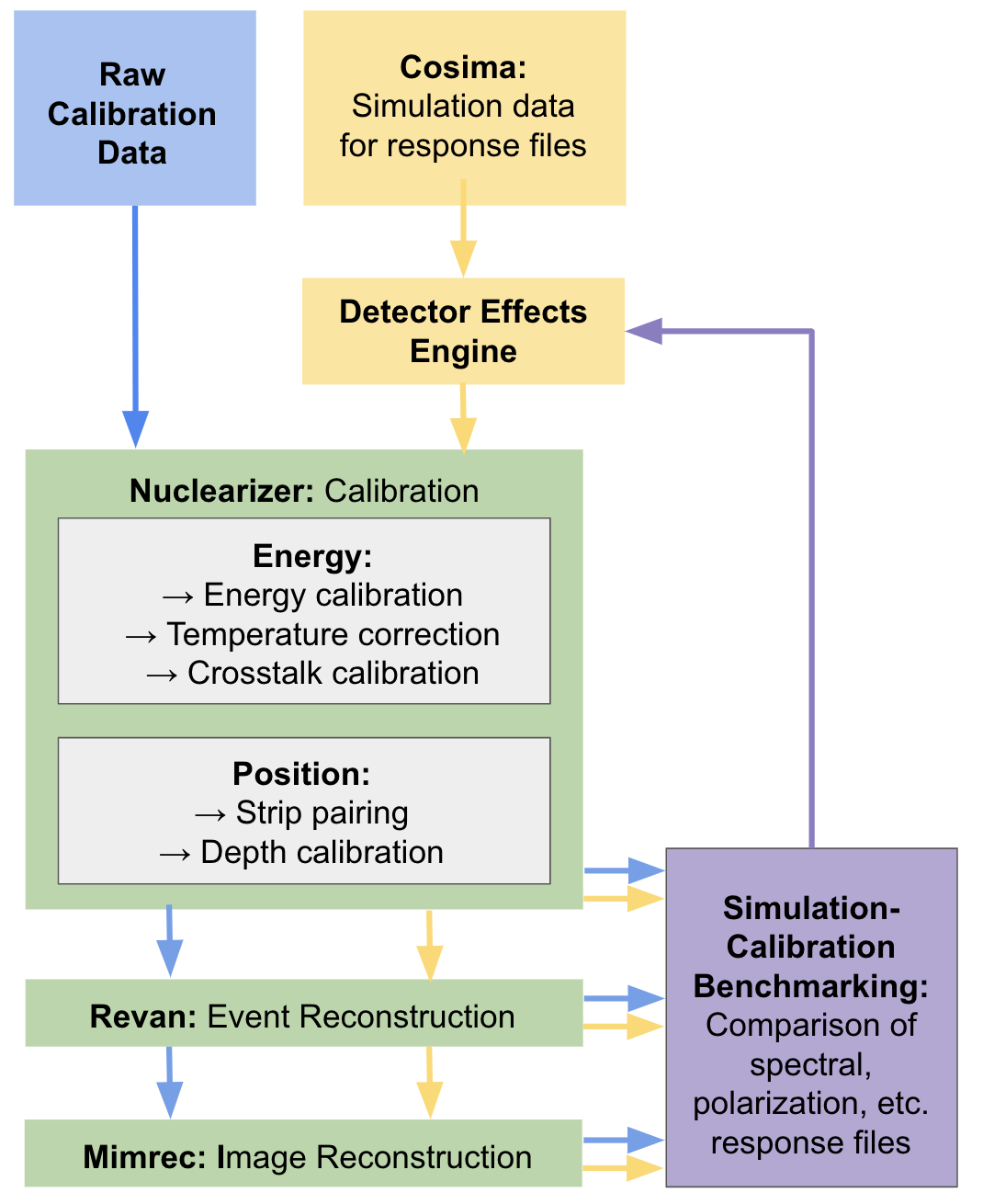}
\caption{COSI data analysis pipeline.}
\label{fig:pipeline}
\end{minipage}
\hfill
\begin{minipage}{0.45\textwidth}
\includegraphics[width = 0.9\textwidth]{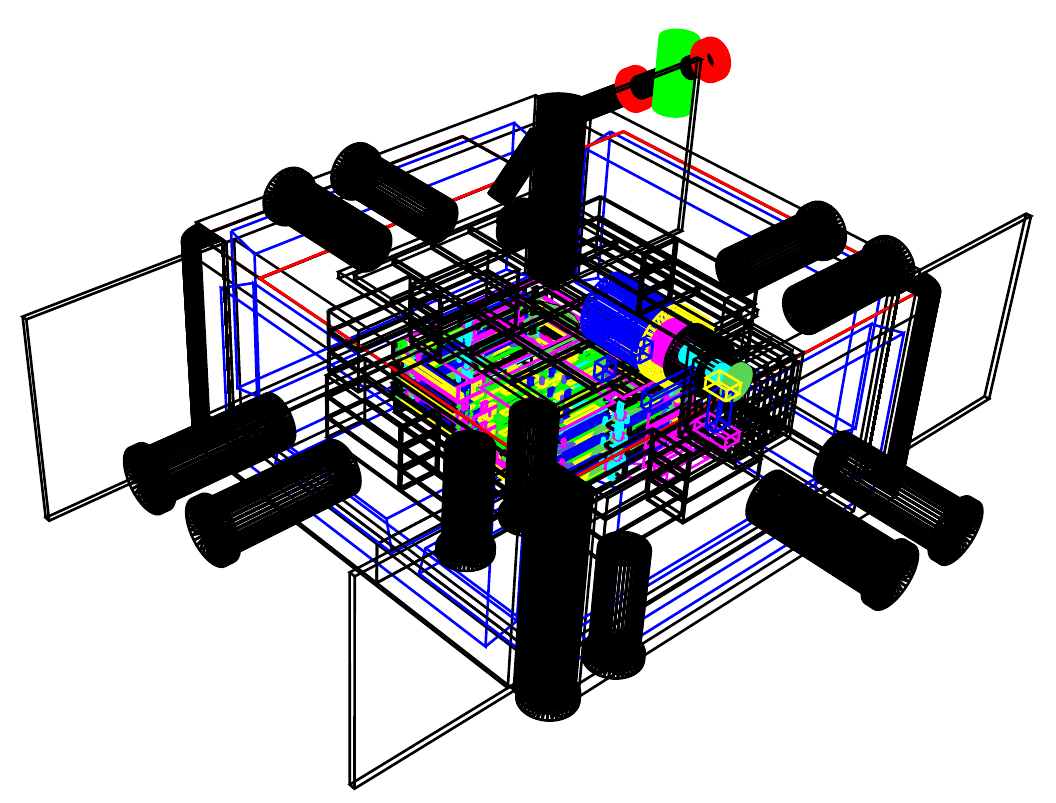}
\caption{Mass model of the COSI detector head made in MEGAlib's Geomega.}
\label{fig:massmodel}
\end{minipage}
\end{figure}

The COSI collaboration uses the MEGAlib software package\footnote{MEGAlib is available at https://github.com/zoglauer/megalib} to perform data analysis and run simulations \cite{zoglauer2006megalib}. This data analysis pipeline, as implemented within the MEGAlib framework, is illustrated in Figure\,\ref{fig:pipeline}. Section\,\ref{sec:calibrations} and Section\,\ref{sec:instrument_performance} further detail each stage of the workflow. MEGAlib provides separate programs for each stage of the data analysis pipeline:

\begin{description}[align=left]
\item [Geomega] (``Geometry for MEGAlib") defines the mass model of the instrument containing detailed information about detector geometry, materials, trigger criteria, and more (Figure\,\ref{fig:massmodel}). 
\item [Cosima] (``Cosmic simulator") generates Monte Carlo simulations of photon interactions within the Geomega instrument mass model. Cosima is an interface to Geant4 \cite{AGOSTINELLI2003250}.
\item [DEE] (``Detector Effects Engine") applies measured detector performance to Cosima simulations such that the simulation data resemble real data taken with the COSI detectors \cite{SLEATOR2019162643} (Section\,\ref{sec:dee}). 
\item [Nuclearizer] performs calibrations which convert measured parameters (charge signal amplitude, strip number) into physical units (energy, position) and applies corrections to real data that rectify imperfections in the measurement process.
\item [Revan] (``Real event analyzer") groups individual hits in simulated and real data into events and performs Compton reconstruction to find the scatter angle of the initial Compton interaction.
\item [Mimrec] (``MEGAlib image reconstruction") employs a list-mode-likelihood iterative scheme to reconstruct images from data. Mimrec can perform high-level data analysis tasks, including studies of energy spectra, angular resolution, polarization, timing distributions, and more.
\end{description}

Real data and simulated data are processed through this same analysis pipeline. Nuclearizer converts measured quantities in real instrument calibration data from electronic units to physical units. When processing simulation data, the DEE converts the pure simulated events, given in physical units, to events mimicking those measured by the detectors, in electronic units. Nuclearizer then converts the DEE output back to physical units of position and energy. This process ensures that simulated data share the imperfections intrinsic to the measurement process seen in real data.

\section{Calibrations}
\label{sec:calibrations}

Calibrations convert measured detector parameters into physical parameters, such as energy and position, necessary for Compton reconstruction. The accuracy of these calibrations is critical to realizing the highest achievable performance of a Compton telescope. 

Nuclearizer calibrates the COSI data in the following steps: 

\begin{enumerate}
  \item \textit{Load data} - The measured data for one event can contain multiple active strips in multiple detectors. For each strip hit we record: Strip ID, Detector ID, ADC, and timing. 
  \item \textit{Energy calibration} - The pulse height, in ADC units, associated with one interaction is converted to a deposited energy in keV (Section\,\ref{sec:energy_calib}).
    \item \textit{Cross-talk correction} - Interactions on neighboring strips  enhance recorded energies. This enhancement, called ``cross-talk," scales linearly with energy and is removed via a linear correction (Section\,\ref{sec:crosstalk}).
  \item \textit{Strip pairing} - The $x$--$y$ position of an interaction in one detector is determined by the intersection of the triggered AC and DC strips. If there is more than one interaction in the detector and multiple strips on each side are triggered, an algorithm determines the most likely interaction position given the energies deposited on all triggered strips (Section\,\ref{sec:strip_pairing}).
  \item \textit{Depth calibration} - The intersection of orthogonal strips is converted into an $x$- and $y$-position in the detector and the difference between electron and hole collection times (the ``collection time difference" or ``CTD") is converted into depth. The depth calibration performs the CTD conversion to depth in physical units (Section\,\ref{sec:depth_calibration}).
  \item \textit{Save calibrated data} - Each event is saved with the energies and positions of its constituent interactions across multiple detectors.
\end{enumerate}

\begin{table}
\centering
\caption{The seven radioactive isotopes used to calibrate COSI. The peak $\gamma$-ray lines of each are listed in keV with their respective branching ratios (BR).}
\begin{tabular}{cc}
Source & Line energy [keV] (BR)    \\      
\hline   
\hline
$\mathrm{^{241}Am}$ & 59.5 (35.9\%)       \\
$\mathrm{^{57}Co}$  & 122.1 (85.6\%), 136.5 (10.7\%)    \\
$\mathrm{^{133}Ba}$ & 81.0 (34.1\%), 276.4 (7.1\%), 302.9 (18.3\%), \\ & 356.0 (62.1\%), 383.85 (8.9\%) \\
$\mathrm{^{22}Na}$  & 511.0 (180.7\%), 1274.5 (99.9\%)    \\
$\mathrm{^{137}Cs}$ & 661.7 (85.1\%)  \\
$\mathrm{^{88}Y}$   & 898.0 (93.7\%), 1836.0 (99.2\%)  \\
$\mathrm{^{60}Co}$  & 1173.2 (99.97\%), 1332.5 (99.99\%) \\
\hline
\end{tabular}
\label{table:isotopes}
\end{table}

\subsection{Data Collection}
\label{sec:data_collection}
To collect calibration data, we use Isotrak Eckert \& Ziegler Type D sealed $\gamma$-ray sources ($\mathrm{^{241}Am}$, $\mathrm{^{57}Co}$, $\mathrm{^{133}Ba}$, $\mathrm{^{22}Na}$, $\mathrm{^{137}Cs}$, $\mathrm{^{88}Y}$, $\mathrm{^{60}Co}$) which yield fifteen nuclear lines within COSI's energy range (Table\,\ref{table:isotopes}). These point-like sources are mounted in a variety of positions surrounding the instrument in order to illuminate the entire field of view. Data were collected in three configurations: low-energy (LE; $<511
$\,keV), high-energy (HE; $\ge511$\,keV), and polarized radiation data collection.

\subsubsection{Low-energy configuration}
Low-energy sources ($\mathrm{^{241}Am}$, $\mathrm{^{57}Co}$, and $\mathrm{^{133}Ba}$) suffer appreciable attenuation and are unable to penetrate the full depth of each GeD, let alone the full COSI GeD stack. As such, each low-energy source is placed in numerous positions in the immediate vicinity of the cryostat until all strips on all detectors have been sufficiently exposed. Collection times in the ten or more positions can range from minutes to hours depending on the activity of the source and the physical accessibility of the strips in question. Given that source activities varied between 2016 and 2020, the exact positions and integration times for the low-energy sources changed but the general approach of moving the sources around the cryostat remained the same.

All low-energy sources are used to conduct a LE energy calibration. In 2016, $\mathrm{^{241}Am}$ data (in conjunction with $\mathrm{^{137}Cs}$ data) were used to perform a temperature correction (Section\,\ref{sec:temp_correction}). The cross-talk corrections in 2016 and 2020 used a range of LE sources together with HE sources described in the next section.

\subsubsection{High-energy configuration}
In 2016, the radioactive sources were held in place using the calibration structure described in \cite{kierans2018}. For the HE energy calibration, sources were placed at a height of $\sim$ 63\,cm above the center of the detector stack for a minimum of five hours per source. Additionally, $\mathrm{^{137}Cs}$ data were collected at the zenith of the 2016 calibration structure for the 2016 temperature correction and depth calibration. The HE sources were held in numerous positions on the calibration structure for angular resolution and effective area studies. These calibration runs spanned several hours each in order to collect data over COSI's entire field of view with ample statistics. These data facilitated studies of COSI's angular resolution and effective area as a function of photon energy and position in the field of view. 

\begin{figure}[htpb]
\centering
\includegraphics[scale=0.07]{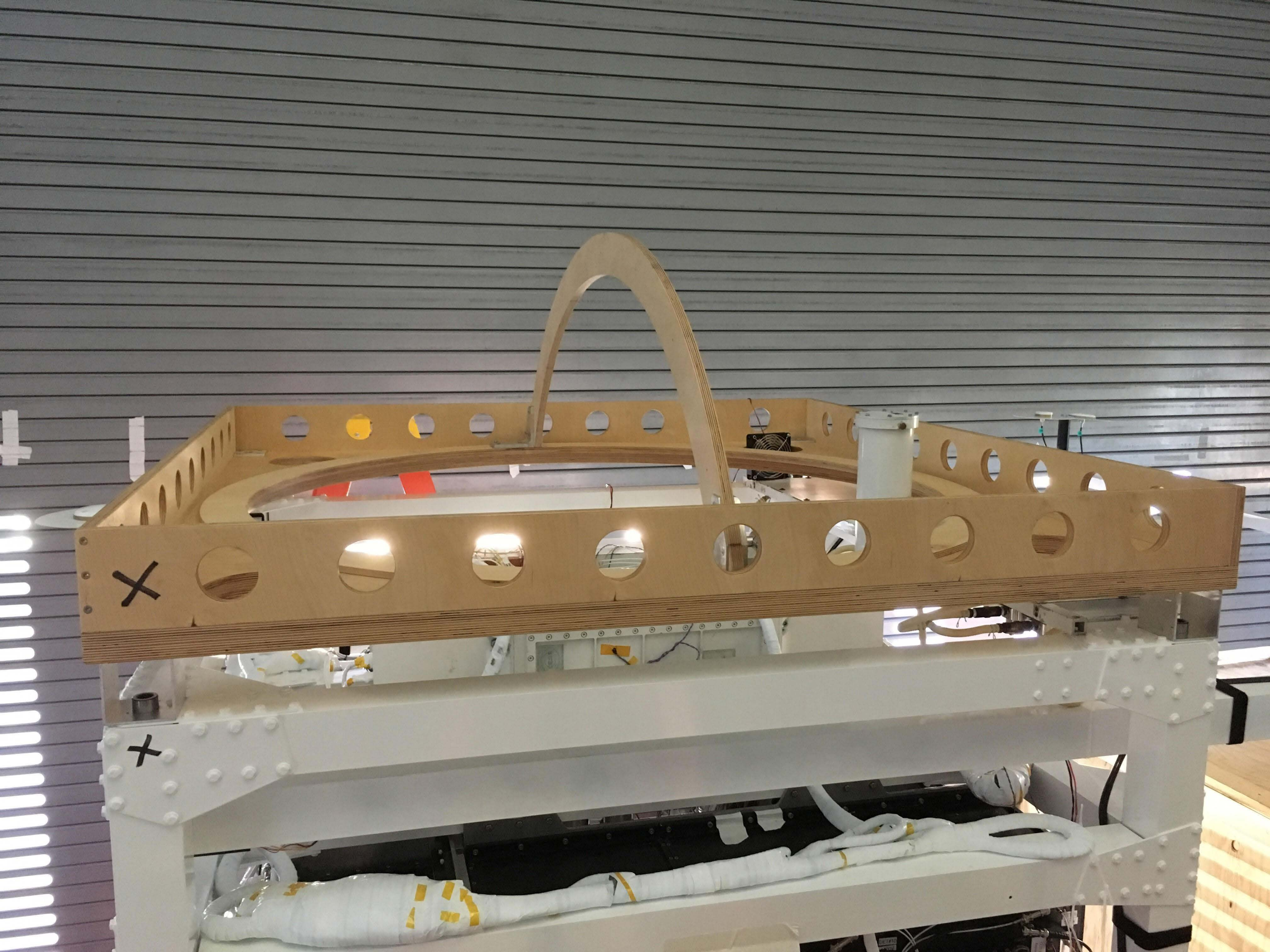}
\caption{The custom-built COSI calibration structure used in 2020 for HE energy calibration, depth calibration, angular resolution, and effective area studies. The plywood structure holds radioactive sources above the cryostat and is shown mounted to the top of the gondola.}
\label{fig:calibration_structure}
\end{figure}

In 2020, a custom-built calibration  structure (Figure\,\ref{fig:calibration_structure}) was used to collect data for HE energy calibration, temperature correction, cross-talk correction, depth calibration, angular resolution, and effective area studies. It was designed to secure sources over COSI's entire field of view in reproducible positions that could be easily mimicked in simulations. The structure was built at the SSL machine shop from plywood and is attached to the top of the gondola with four bolts on each corner, one of which is visible between the ``X" marks in the bottom left corner of Figure\,\ref{fig:calibration_structure}. It is only affixed to the top of the gondola when collecting calibration data and is mounted onto and lifted off the gondola by hand.

Radioactive sources are secured to the protruding arch with a Delrin plastic source holder that is tightened with a nylon screw to any polar angle along the arch. The arch rotates freely in the azimuthal direction and the polar angle spans 0$^{\circ}$ at zenith to approximately 60$^{\circ}$ on either side, enabling complete illumination of COSI's field of view. The radius of the arch is $\sim$ 63\,cm. Moving the sources to various zenith angles along the arch and rotating azimuthally characterizes COSI's performance over the entire field of view. 

In the 2020 HE energy calibration, the COSI team placed $\mathrm{^{137}Cs}$, $\mathrm{^{60}Co}$, and $\mathrm{^{88}Y}$ at zenith but closer to the cryostat ($\sim$14.6\,cm, $\sim$14.6\,cm, and resting atop, respectively) for expedited data collection. $\mathrm{^{22}Na}$ data were collected at the zenith of the wooden calibration structure. These data runs spanned at least 7 and up to 24 hours. 

The temperature correction in 2020 (Section\,\ref{sec:temp_correction}) used $\mathrm{^{22}Na}$ data collected from the zenith of the calibration structure. The depth calibration used $\mathrm{^{137}Cs}$ data taken from the zenith of the calibration structure. Angular resolution and effective area data were also collected using the calibration structure, but limited time for calibrations prevented the team from collecting data over the entire field of view as in 2016. In 2020, the angular resolution and effective area measurements were limited to $\mathrm{^{60}Co}$, $\mathrm{^{137}Cs}$, and $\mathrm{^{22}Na}$ at the zenith of the calibration structure. As in 2016, these data runs spanned several hours.

\subsubsection{Polarized radiation collection}

\begin{figure}[htpb]
\begin{subfigure}{.5\textwidth}
  \centering
  \includegraphics[width=.94\linewidth]{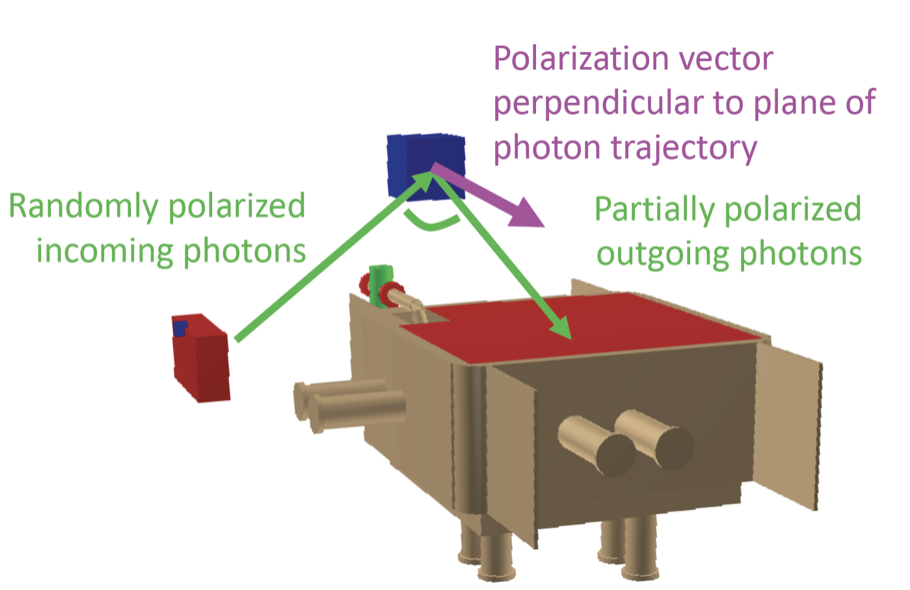}  
  \caption{}
  \label{fig:model}
\end{subfigure}
\begin{subfigure}{.48\textwidth}
  \centering
  \includegraphics[width=.9\linewidth]{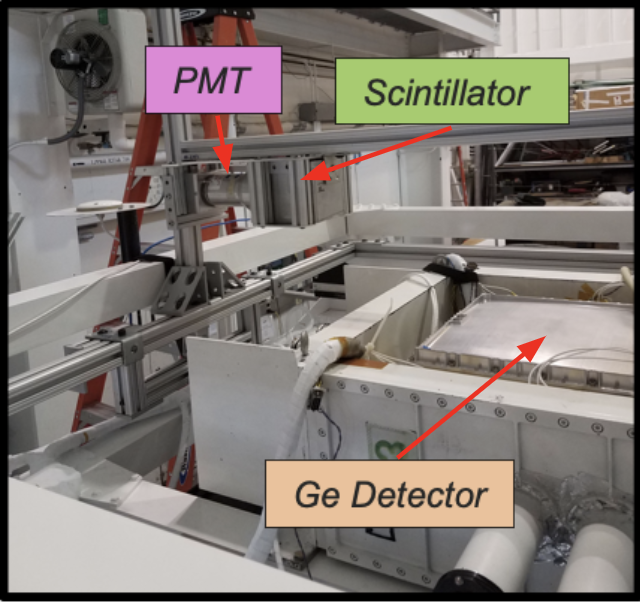}  
  \caption{}
  \label{fig:polar_lab}
\end{subfigure}
\caption{(a) To produce a polarized beam, $\gamma$-rays are emitted from a source and scatter off a scintillator towards the detector. (b) The 2020 campaign polarization calibration structure.}
\label{fig:polar}
\end{figure}

We are able to produce partially-polarized $\gamma$-rays using a principle detailed in \cite{lei97}. When unpolarized photons Compton scatter, the outgoing beam is partially polarized with a polarization level of $\Pi$, given by

\begin{equation}
    \Pi = \frac{\sin^{2} \theta}{\epsilon + \epsilon^{-1} - \sin^{2}\theta},
\label{eq:polarization}
\end{equation}

\noindent in which $\epsilon$ is the ratio of scattered photon energy to initial photon energy and $\theta$ is the scattering angle. The polarization vector of the scattered beam is perpendicular to the scattering plane, as seen in Figure\,\ref{fig:model}. We produced partially-polarized $\gamma$-ray beams in the laboratory by scattering photons from a Sodium Iodide (NaI) scintillator, which has an attached PMT. The time stamps of events recorded by the PMT allow us to select only events coincident between COSI and the NaI. As the count rate of the scattered photons is low ($<$ 1 count s$^{-1}$), this coincidence technique was designed for its ability to reject the majority of the background. 

\begin{figure}[htpb]
\centering
\includegraphics[width = 0.7\textwidth]{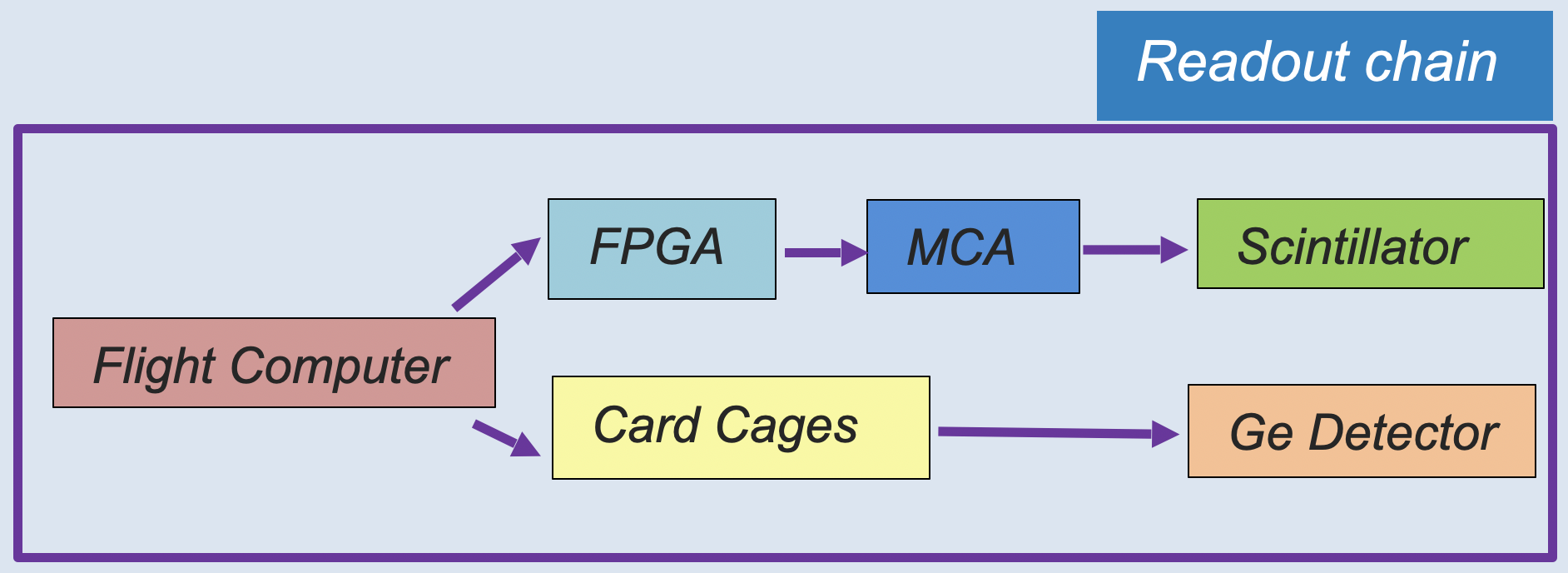}
\caption{Readout chain to synchronize clock values of the NaI and Ge detectors.}
\label{fig:sync}
\end{figure}

\begin{figure}[htpb]
\centering
\includegraphics[width = 0.8\textwidth]{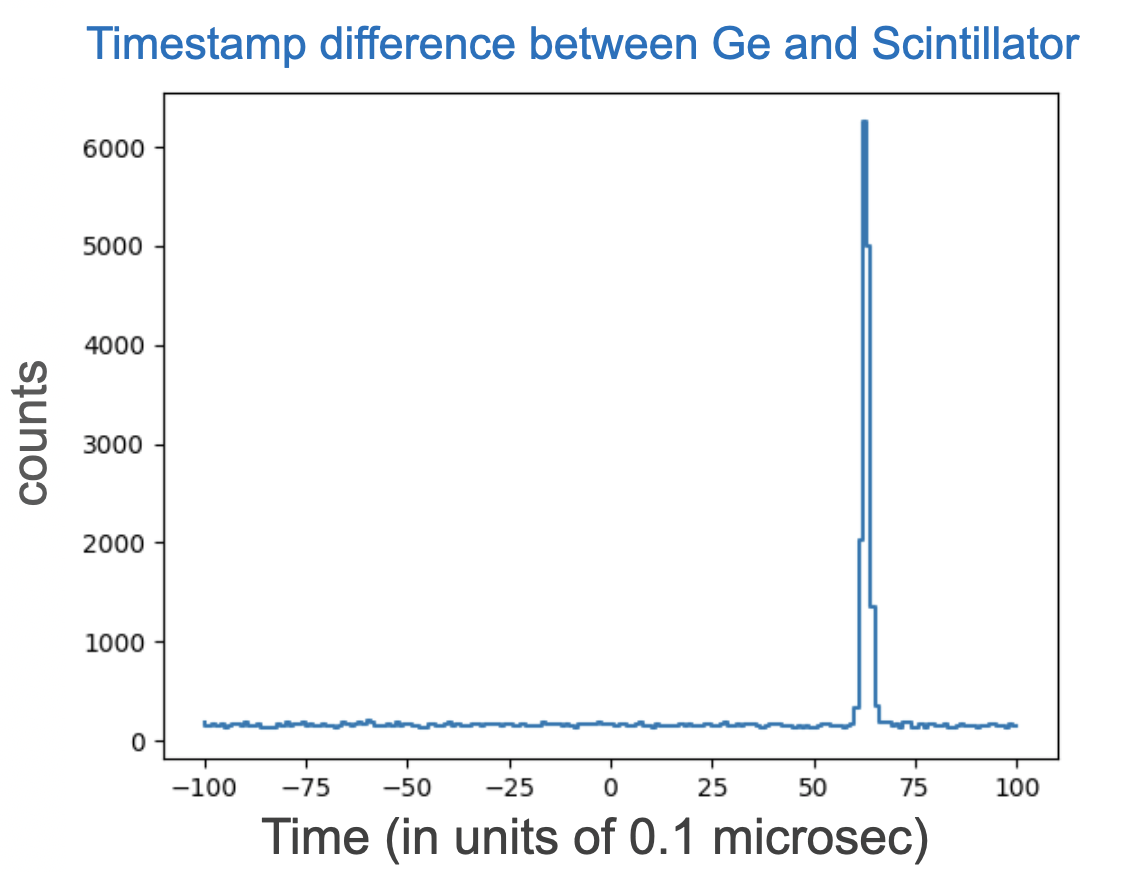}
\caption{A time difference histogram, where the $x$-axis represents the time between a COSI event and the time of the next NaI time stamp. A clear coincidence peak is seen between 6000\,ns and 7000\,ns. These coincident events are selected to isolate the partially-polarized beam.}
\label{fig:time_diff}
\end{figure}

The PMT was attached to an MCA board with a detector for pulse-height measurements and an FPGA that synchronized the time stamps of these peaks to those of the card cages. The readout chain was initiated with a ``sync" signal sent by the flight computer's 10\,MHz oscillator, as seen in Figure\,\ref{fig:sync}. This pulse was delivered to the sync line of the card cages and that of the FPGA. The lag time between the NaI scintillator and Ge is determined by the spread in the time difference between the PMT  and card cages, respectively (see Figure\,\ref{fig:time_diff}). Thus the spread provided a $\pm$6\,$\mu$s time window for coincident events for the NaI and Ge.

The 2020 campaign was canceled due to restrictions arising from the COVID-19 pandemic. While all other calibration data were collected at the planned launch site, the cancellation prevented polarization response data collection. Thus the configuration for the 2019 pre-flight data acquisition test at SSL is provided as an example in Figure\,\ref{fig:polar_lab}. A custom-built calibration structure for the 2020 campaign suspends the NaI detector above the cryostat and fixes a $\mathrm{^{137}Cs}$ source at the cryostat’s level. It also holds a lead brick between the $\mathrm{^{137}Cs}$ source and the COSI detector system to prevent the direct flux from unnecessarily elevating the shield count rate and thereby vetoing desired events. The apparatus captures a number of Compton scattering angles, defined by $\theta$ in Equation\,\ref{eq:polarization}, as well as polarization angles, defined by the angle of the plane of the photon trajectory.

\subsection{Energy Calibration}
\label{sec:energy_calib}

Each of COSI's strips is read out individually by the data acquisition system and is uniquely calibrated. The energy calibration for each strip is determined by collecting data from radioactive sources with known $\gamma$-ray line energies (Table\,\ref{table:isotopes}) and measuring the resulting pulse height signal in the electronics. Identifying multiple lines across COSI's energy range and fitting with a polynomial yields the desired relationship between pulse height and energy for each strip. In this section we detail the analysis procedures and results from energy calibrations performed before the Wanaka 2016 launch and before the intended launch from Wanaka in 2020. These calibrations define the single-strip spectral resolution of COSI.

Energy calibration analysis uses the software ``Melinator" (``MEGAlib's line calibrator"). The summed, raw spectra of all collected data are loaded into Melinator one detector at a time to reduce computational strain. Melinator fits the photopeaks seen on each strip in ADC space (the pulse heights) with a Gaussian (convolved with a delta function for energy loss) and a linear background model. The fitting algorithm returns the centroid value of each fitted peak in ADC and matches it with the corresponding known, true photopeak energy in keV (Table\,\ref{table:isotopes}).

\begin{figure}[htbp]
\centering
\includegraphics[scale=0.33]{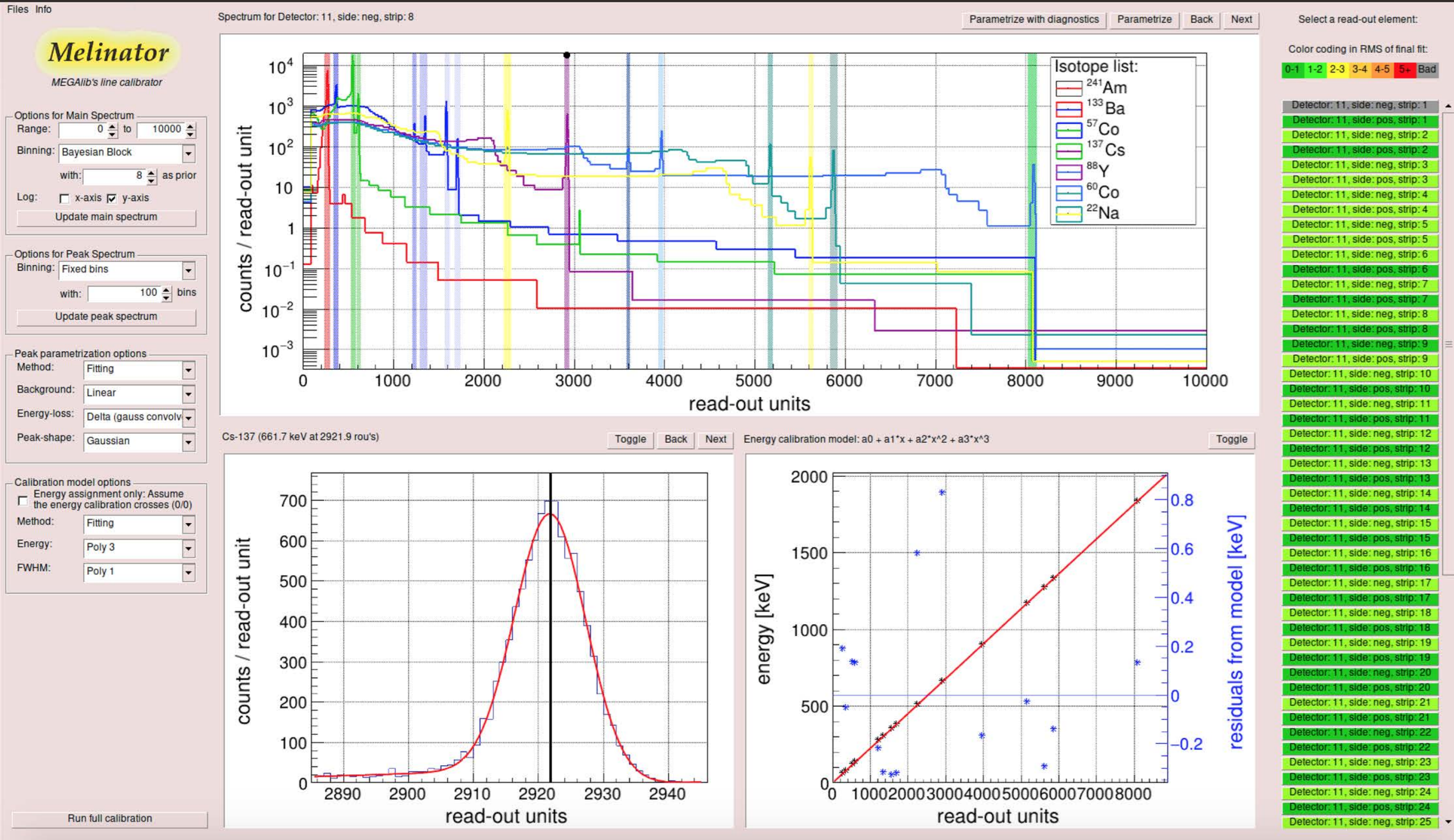}
\caption{An example of the 2020 energy calibration for strip 8 on the DC side of detector 11, as performed by Melinator. Melinator identifies the photopeaks of all seven isotopes on each strip's cumulative spectrum (top center plot) and fits a third-order polynomial to the resulting scatter plot of energy [keV] vs. read-out units [ADC] (bottom right plot). The $\mathrm{^{137}Cs}$ 661.7\,keV photopeak on this strip, identified at 2921.9 read-out units, is shown in the bottom left plot.}
\label{fig:melinatorspectra}
\end{figure}

We fit the energy versus ADC relation for each strip with a third-order polynomial to account for non-linearities at low energies. Refer to Figure\,\ref{fig:melinatorspectra} for a screen capture of the Melinator window in the 2020 energy calibration analysis. Melinator also returns the full-width half maximum (FWHM) of each peak in keV. The third-order polynomial defines the conversion of all interaction energies from electronic to physical units and the FWHM is the primary metric of COSI's single-strip, energy-dependent spectral resolution.

Differences in energy calibrations of the instrument are expected and motivate repeated calibrations before each campaign. Electronic repairs and gain adjustments designed to improve performance of analog boards, for example, are a routine part of detector maintenance that change the ADC-energy relationship on affected strips. 

We compare the FWHMs from Melinator in 2016 and 2020. The spectral resolution of the telescope's strips is given as the ratio of the FWHM of the $\mathrm{^{137}Cs}$ $\gamma$-ray line to its photopeak energy of 661.7\,keV. By this definition, the single-strip spectral resolution of COSI in 2016 was $0.453 \pm 0.004$\% on the AC side and $0.45 \pm 0.01$\% on the DC side. In 2020, it was $0.52 \pm 0.01$\% on the AC side and $0.48 \pm 0.01$\% on the DC side (Table\,\ref{table:FWHM_table}).

\begin{figure}[htbp]
\centering
\includegraphics[scale=0.5]{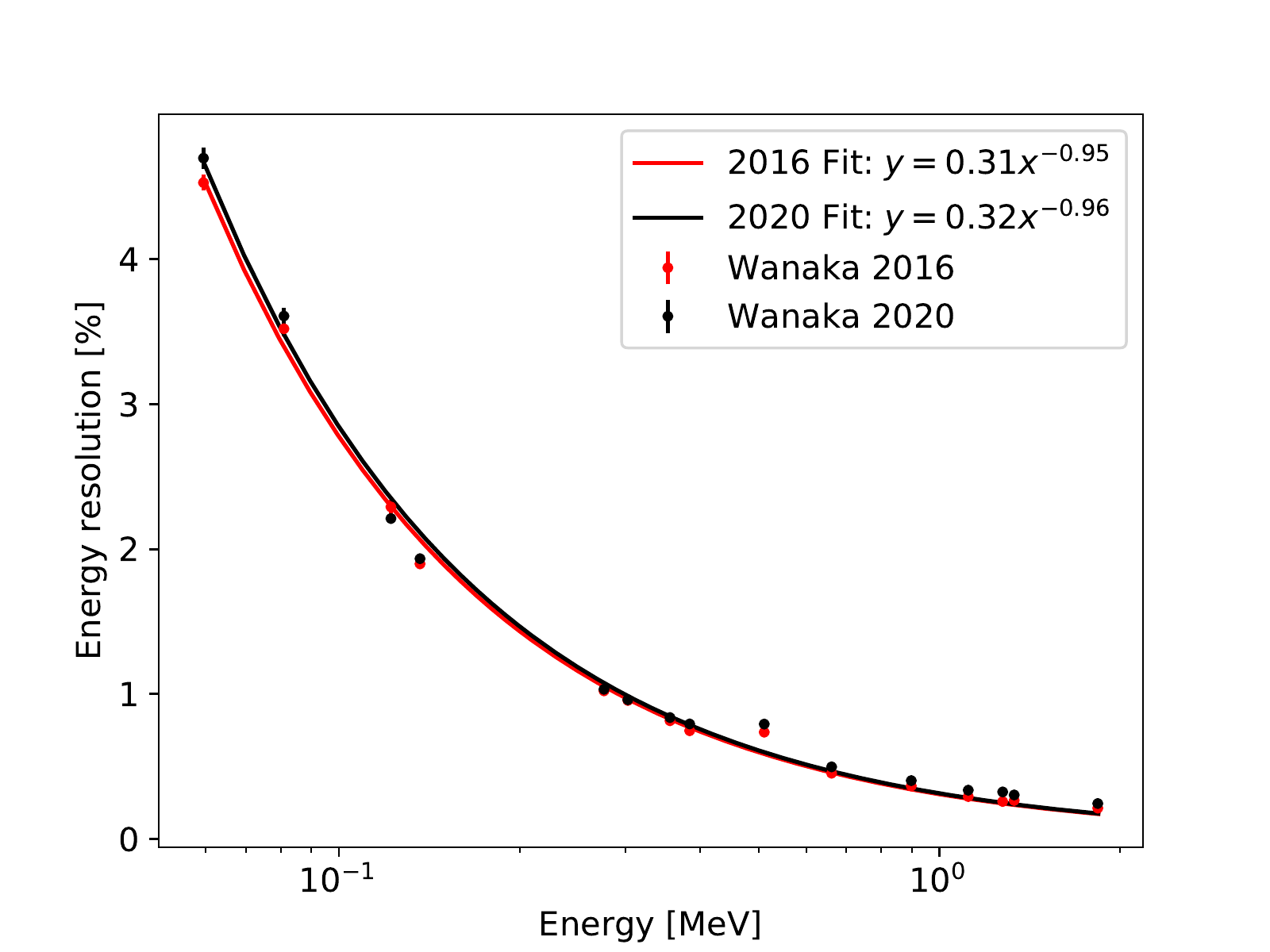}
\caption{Single-strip energy resolution (FWHM) as a function of energy in 2020 and 2016 energy calibrations. The fitted power law exponent $k = -0.96$ is consistent with the $1/N$ dependence expected from dominant electronic noise in COSI's detectors.}
\label{fig:FWHMvsenergy}
\end{figure}

Figure \ref{fig:FWHMvsenergy} demonstrates the dependence of COSI's single-strip energy resolution on photon energy $E$. The observed dependence aligns with expectations of dominant electronic noise in COSI up to about 1\,MeV. Electronic noise scales as $1/N$, where $N$ is the number of charge carriers generated in an interaction. As $N = E/W$ for $W$ = 2.96\,eV, the energy required to produce an electron-hole pair in germanium, we expect energy resolution $\propto 1/E$. Indeed, fitting the data points of energy resolution versus energy with a power law gives a power law index $k = -0.96$, matching expectations derived from electronic noise.

\begin{table}[htbp]
\centering
\caption{Mean single-strip energy resolution of COSI's AC and DC strips in 2016 and 2020. The resolution is defined as the ratio of the FWHM of the $\mathrm{^{137}Cs}$ photopeak to 661.7\,keV.}
\resizebox{\textwidth}{!}{
\begin{tabular}{ccccc}
Strips, energy & 2020 FWHM [\%] & 2020 FWHM [keV] & 2016 FWHM [\%] & 2016 FWHM [keV]  \\
\hline
\hline
AC, 662 keV & $0.52 \pm 0.01$  &  $3.42 \pm 0.06$ & $0.453 \pm 0.004$  & $3.00 \pm 0.03$  \\
DC, 662 keV & $0.48 \pm 0.01$ &  $3.17 \pm 0.04$ & $0.45 \pm 0.01$  & $3.00 \pm 0.05$ \\
\hline
\end{tabular}
}
\label{table:FWHM_table}
\end{table}

\subsection{Temperature dependence}
\label{sec:temp_correction}
COSI's preamplifier boards exhibit a temperature dependence which shifts the pulse-height spectra of calibration data and manifests as photopeak energies displaced from their true photopeak energies. The temperature sensitivity was observed to shift spectra by up to 0.5\,keV/$^\circ$C at 661.7\,keV \cite{kierans2018}. Correcting this shift is necessary before proceeding with the subsequently described calibration steps and analysis of instrument performance (Section\,\ref{sec:instrument_performance}) because these steps and analyses rely on accurate energy determination.

To correct the 2016 spectra, $\mathrm{^{137}Cs}$ and $\mathrm{^{241}Am}$ energy calibration data were collected over a wide range of temperatures, approximately 12$^\circ$C to 34$^\circ$C, meant to mimic temperatures seen during flight. 
A linear relationship between preamplifier temperature and ADC peak location was determined for each strip, yielding a precise correction tailored to each strip's individual readout. Before applying the correction, the difference between measured and true line energy was 0.5\%. Applying the correction limited this discrepancy to 0.1\% \cite{kierans2018}.

Unfortunately, the 2020 mission was canceled before the COSI team could take calibration data over the wide temperature range required for the 2016 temperature correction method. 
An alternate method of temperature correction was developed for 2020 data which also resulted in an average offset of 0.1\%.
Further investigation of the temperature dependence is underway, including efforts to reconcile the two methods and encode a potential energy dependence in the temperature-induced shift. We emphasize the importance of characterizing temperature dependence in detector readout and stress that dedicated calibration time at controlled temperatures would greatly benefit this effort.

\subsection{Cross-talk Correction}
\label{sec:crosstalk}

\begin{figure}[htpb]
\centering
\begin{minipage}{0.4\textwidth}
\centering
\includegraphics[height=4.4cm]{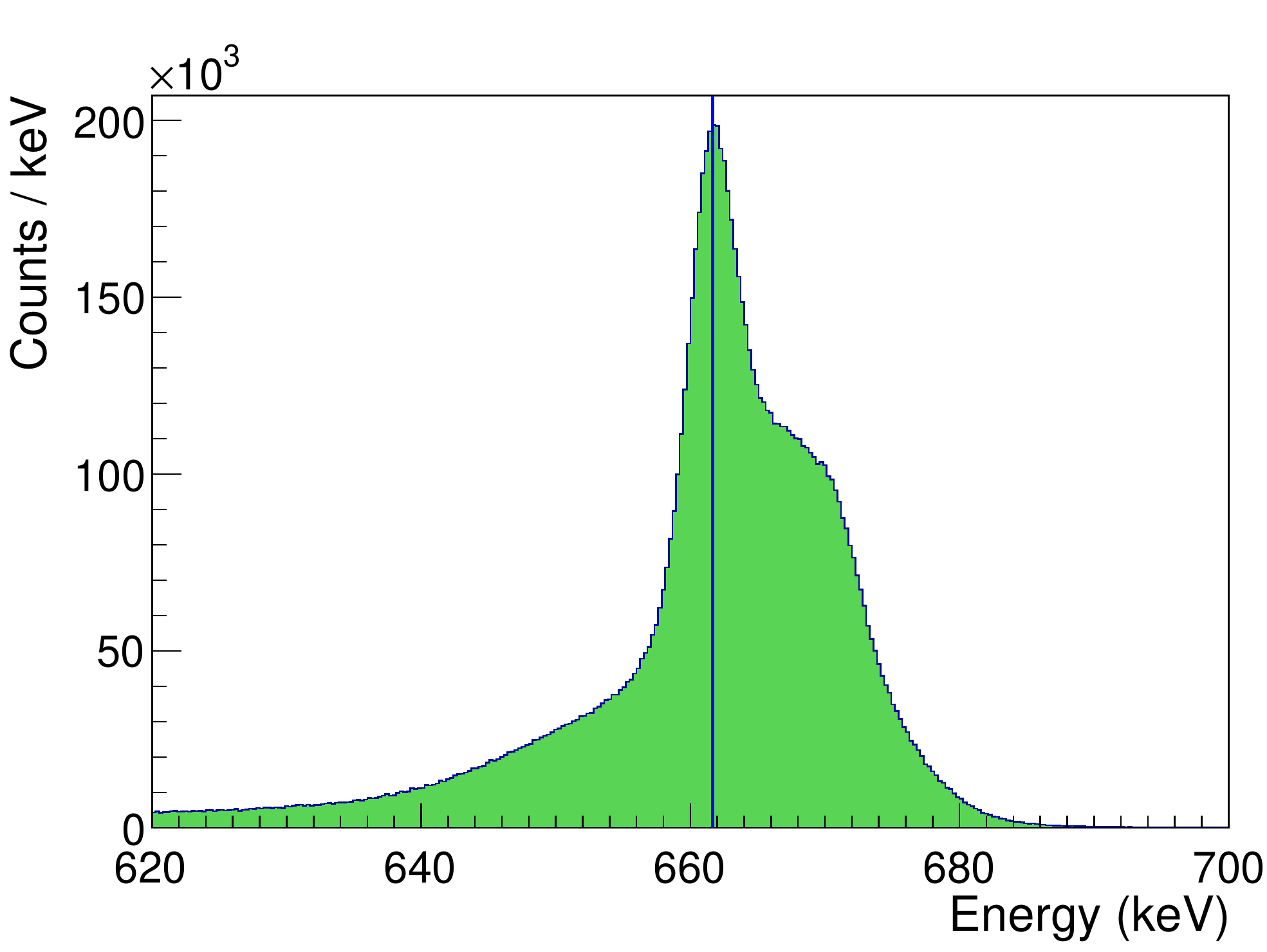}
\caption{Energy spectrum of $\mathrm{^{137}Cs}$ data without correcting for cross-talk. Image from \cite{kierans2018}.}
\label{fig:crosstalkbefore}
\end{minipage}\hfill
\begin{minipage}{0.5\textwidth}
\centering
\includegraphics[height=4.1cm]{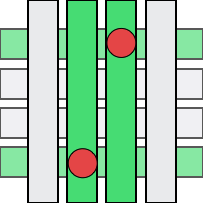}
\caption{Schematic of a multi-site event in which two neighboring strips are triggered.}
\label{fig:crosstalkschematic}
\end{minipage}
\end{figure}

The energy recorded by a strip electrode in close proximity to another triggered strip is enhanced by the charge deposited on the neighboring strip. Thus, without correcting for the induced enhancement, the recorded energies for these events will be higher than the energies deposited by the photons. This means that for a fully reconstructed Compton spectrum of calibration data, there will be an additional feature visible past the line. A visualization of the effects of cross-talk in $\mathrm{^{137}Cs}$ data is shown in Figure\,\ref{fig:crosstalkbefore}. We expect a single Gaussian peak at $\sim$662\,keV, yet there is an additional enhancement at $\sim$670\,keV caused by the influence of neighboring strips. 

\begin{figure}
\centering
    \begin{subfigure}[t]{0.47\textwidth}
         \centering
         \includegraphics[width=0.8\textwidth]{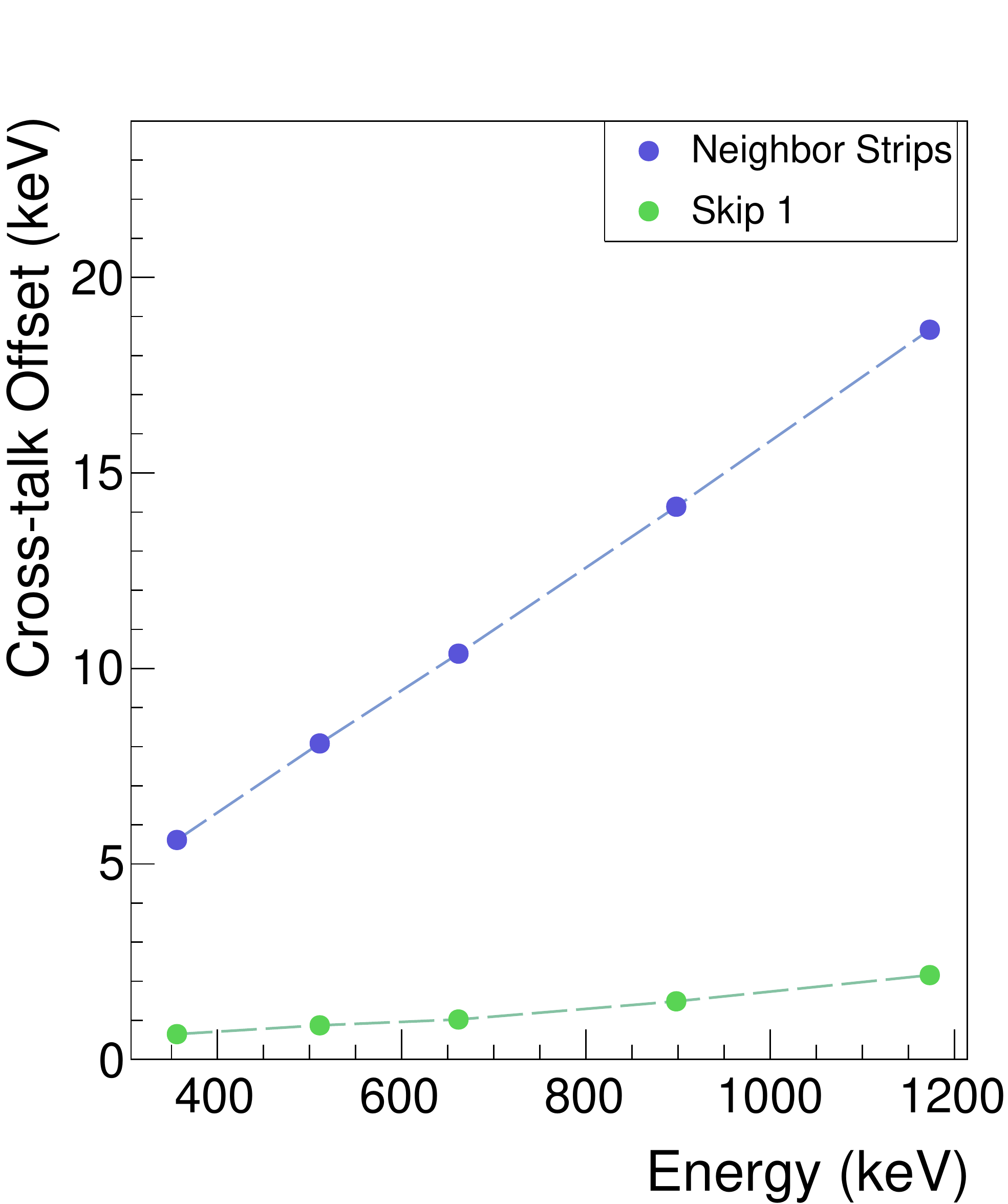}
         \caption{Cross-talk offset versus energy. Image from \cite{kierans2018}.}
         \label{fig:crosstalk}
         \end{subfigure}\hfill
     \begin{subfigure}[t]{0.47\textwidth}
         \centering
         \includegraphics[width=\textwidth]{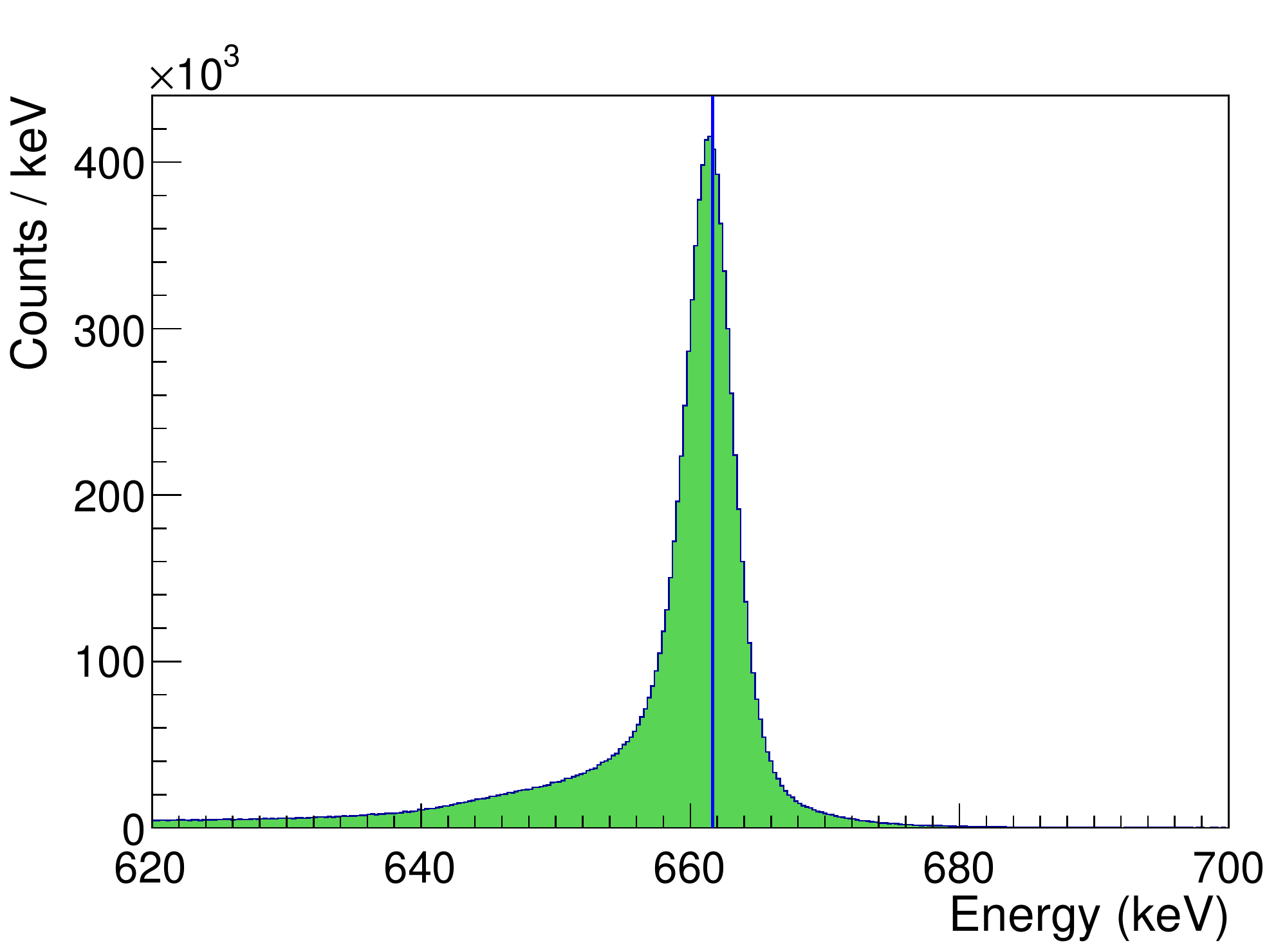}
         \caption{Energy spectrum of $\mathrm{^{137}Cs}$ after applying the linear cross-talk correction. Image from \cite{kierans2018}.}
         \label{fig:crosstalkafter}
     \end{subfigure}
\caption{The linear cross-talk correction. (a) The cross-talk offset scales linearly with the energy recorded by nearest-neighbor and ``Skip 1" strips. (b) Applying this linear correction to $\mathrm{^{137}Cs}$ data removes the enhancement in energy seen in the uncorrected spectrum (Figure\,\ref{fig:crosstalkbefore}).}
\end{figure}

 We determine the cross-talk correction factor by isolating events that have two activated neighboring strips on one side (Figure\,\ref{fig:crosstalkschematic}), as well as next-nearest-neighboring strips, referred to as ``Skip 1." Considering only these events, we determine the energy enhancement as a linear function of line energy, and thus this correction factor is determined by the slope (= cross-talk offset/line energy). We determine one linear correction for each side of each detector for neighboring and Skip 1 events. Figure\,\ref{fig:crosstalk} shows the measured enhancement and linear correction for the DC-side of Detector 6, and Figure\,\ref{fig:crosstalkafter} shows the same spectrum as Figure\,\ref{fig:crosstalkbefore} but with the cross-talk correction included. The average cross-talk correction factors for the AC and DC side strips across the twelve detectors are consistent between the 2016 and 2020 campaigns, as seen in Table\,\ref{table:cross}.

\begin{table}
\centering
\caption{The average cross-talk correction factors for nearest neighboring (NN) and Skip 1 events across the 12 detectors.}
\begin{tabular}{ccc}
 & 2016 & 2020    \\
\hline   
\hline
AC NN & $0.017 \pm 0.001$ &  $0.017 \pm 0.003$     \\
DC NN & $0.0152 \pm 0.0004$ & $0.015 \pm 0.001$       \\
\hline
AC Skip 1 & $0.003 \pm 0.001$ &  $0.004 \pm 0.002$     \\
DC Skip 1 & $0.0023 \pm 0.0002$ & $0.003 \pm 0.001$       \\
\hline
\end{tabular}
\label{table:cross}
\end{table}

\subsection{Strip Pairing}
\label{sec:strip_pairing}

\begin{figure}
\centering
    \begin{subfigure}{0.48\textwidth}
         \centering
         \includegraphics[width=\textwidth]{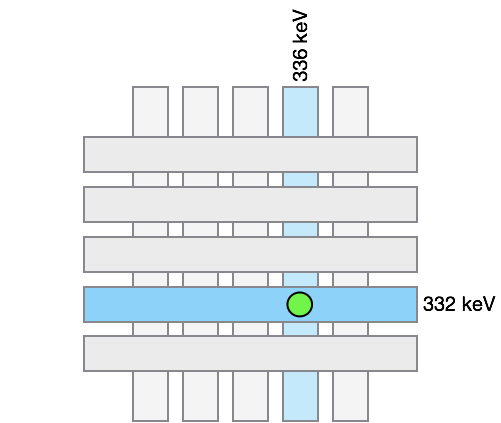}
         \caption{One interaction in a single detector}
         \label{fig:stripPairing1Hit}
     \end{subfigure}
     \begin{subfigure}{0.48\textwidth}
         \centering
         \includegraphics[width=\textwidth]{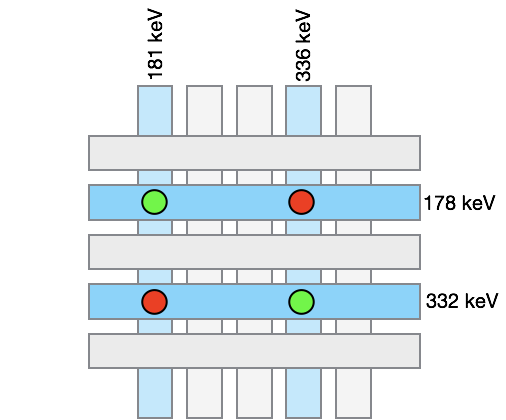}
         \caption{Two interactions in a single detector}
         \label{fig:stripPairing2Hits}
     \end{subfigure}
\caption{For strip pairing (a) with one interaction in a detector, determining the $x$--$y$ position -- the position where the AC and DC side strips overlap -- is straightforward. (b) With two interactions in a detector, there are two possible solutions. We determine the correct solution, marked here by the green circles, by comparing the energies deposited on the strips. Figure is from \cite{sleator2019measuring}.}
\end{figure}

To determine the $x$--$y$ position of an interaction, we use the triggered AC and DC side strip IDs in a process known as strip pairing. Strip pairing is described in detail in \cite{sleator2019measuring} and is summarized here. If there is only one interaction in a detector, the process is straightforward: the $x$--$y$ position is the point at which the AC and DC side strips intersect (Figure\,\ref{fig:stripPairing1Hit}). If there are multiple interactions in the detector, the process becomes more complicated since there are several candidate interaction locations. Figure\,\ref{fig:stripPairing2Hits} shows a schematic of the case in which two interactions occur in a detector, which can lead to two solutions, marked by the green and red circles. As the number of interactions increases, so does the number of possible solutions.

To determine the correct solution, we compare the energy deposited on each strip hit. In the example shown in Figure\,\ref{fig:stripPairing2Hits}, the green circles represent the correct solution because the energies deposited on the AC and DC side strips of each pair match more closely. To perform the strip pairing, we use a Greedy (e.g. \cite{jungnickel2005greedy}) algorithm that compares each AC strip energy to each DC strip energy and finds the closest match. Greedy algorithms choose the locally optimal choice without considering the global consequences of that choice, and are advantageous because they approximate the optimal solution with fewer computation steps. The strips that constitute the closest match are removed from the pool and the algorithm then chooses the pair with the closest match in energies from the remaining strips. The algorithm continues choosing and removing the best pair among the available strips from the pool until all of the strips are paired.

The closest match is determined by the local quality factor $Q_{i,j}$:

\begin{equation}
    Q_{i,j} = \frac{(E_i^{\text{AC}}-E_j^{\text{DC}})^2}{(\sigma_i^{\text{AC}})^2+(\sigma_j^{\text{DC}})^2}
\end{equation}

\noindent where $i$ and $j$ denote the specific strips on the AC and DC sides, $E$ is the energy deposited on the strip, and $\sigma$ is the energy resolution of that strip. $Q_{i,j}$ compares the agreement of the AC and DC side energies within measurement uncertainties: the lower the $Q_{i,j}$, the better the agreement.

The strip pairing process is further complicated by detector and electronics effects such as finite energy resolution, charge sharing between adjacent strips, multiple interactions occurring on a single strip, and energy distortion due to charge loss, including sub-threshold energy deposits (see \cite{SLEATOR2019162643} for a detailed description of these effects). To assess the success of the strip pairing algorithm, we calculate a global quality factor $Q$, which is the sum of the local quality factors of each of $N$ total pairs:

\begin{equation}
    Q = \frac{1}{N}\sum^{N}_{i,j=1}Q_{i,j} = \sum^{N}_{i,j=1}\frac{(E_i^{\text{AC}}-E_j^{\text{DC}})^2}{(\sigma_i^{\text{AC}})^2+(\sigma_j^{\text{DC}})^2}
\end{equation}

If $Q$ is greater than the maximum acceptable value $Q_{\text{max}}=25$, the event is deemed too complicated to pair, likely due to a combination of the detector effects discussed above. Events with $Q>Q_{\text{max}}$ are flagged to be rejected during later analysis. The fraction of these events is energy dependent and ranges from $\sim17\%$ at 511\,keV to $\sim28\%$ at 1274\,keV \cite{sleator2019measuring}.

\subsection{Depth Calibration}
\label{sec:depth_calibration}
The intersection of orthogonal strips on opposite sides of the GeDs yields the two-dimensional $x$--$y$ position of an interaction. The third dimension, depth, is derived from the collection time difference (CTD, denoted below by $\tau$) of electrons and holes generated in the interaction.

When a charge cloud is generated in the active volume of a GeD, electrons and holes drift in opposite directions along field lines to the orthogonally oriented strips on the AC and DC sides of the detector, respectively. If the charge cloud is produced closer to the AC side of the detector, the collection time of the electron will be less than that of the hole. Hence, the CTD is a proxy for localizing the depth of the interaction. The process of extracting this depth from the CTD is called depth calibration.

\subsubsection{Classic Depth Calibration}
Each of the ``pixels" in the GeDs (regions segmented by the grid of orthogonal strips) is uniquely calibrated to account for individual strip read out and variations in drift velocities across the detector volume. We follow a classic approach to calibrate the 37 strips $\times$ 37 strips $\times$ 12 detectors = 16,428 pixels. The name ``classic" is in reference to its successful implementation in NCT \cite{bandstra:2007} and the ongoing work to improve the calibration for future applications. The outline of this approach is explained in detail in \cite{lowell:2016} and \cite{lowell:phdthesis} and is adapted here for clarity. There are three main ingredients:

\begin{enumerate}
    \item Charge transport simulations 
    
    Charge transport simulations relate the CTD $\tau_{\rm sim}$ to depth $z$ through a look-up table of depth $z_n({\tau_{\rm sim}})$ for each detector $n$. The look-up table is generated by simulations that solve Poisson's equation for the electrostatic potential inside the active volume of a simplified $5 \times 5$ strip GeD given appropriate boundary conditions (detector bias) and detector characteristics (impurity concentrations and thickness). The weighting field \cite{knoll2010radiation} is calculated similarly with different boundary conditions: instead of applying the usual detector bias of 1000--1500\,V to the AC strips, each strip is set to 1\,V while the rest are set to 0\,V. With the weighting field we use the Shockley-Ramo theorem \cite{shockley1938currents} to calculate the current induced on the electrodes by moving charge carriers.
    
    \item Real calibration run 
    
    We collect data with a $\sim$0.08\,mCi $\mathrm{^{137}Cs}$ calibration source placed at the zenith of the calibration structure. Using photopeak (650--670\,keV) and continuum (200--477\,keV) events, a CTD is recorded for each pixel. We collect several hours of data which yield at least one hundred counts in each bin of the per-pixel CTDs (Figure\,\ref{fig:depth_CTDfit}).
    
    \item Simulated calibration run 
    
    A Cosima simulation mimicking the experimental conditions above produces a histogram of interaction depths, i.e. a simulated depth distribution, for each detector. Note that the simulation does not include background because the calibration source is close enough to the detectors such that the spectra are source-dominated.
\end{enumerate}

We convert the simulated depth distributions from Step 3, in units of centimeters, to simulated CTD distributions, in units of nanoseconds, using the twelve look-up tables in Step 1. This simulated CTD distribution for each detector, called a ``CTD template," is then used to calibrate the detector's constituent pixels. 

We fit for the ``stretching" $\lambda$ and ``offset" $\Delta$ factors that most closely transform measured CTDs from each pixel in Step 2 into the CTD template of the detector corresponding to the pixel of interest. See Figure\,\ref{fig:depth_CTDfit} for an example of a successful fit. The transformation is given by 

\begin{equation}
\tau_{\rm meas} = \lambda\tau_{\rm sim} + \Delta
\end{equation}

\noindent and by using the $\lambda$ and $\Delta$ returned for each pixel, we calculate a per-pixel $\tau_{\rm sim}$ which is converted to a depth via the look-up table in Step 1. Thus, we obtain a pixel-specific depth calibration relating the CTD and $z$-coordinate of each interaction. 

\subsubsection{Comparison to 2016}
\begin{figure*}[htbp]
    \centering
    \begin{subfigure}{0.49\textwidth}
            \includegraphics[width=0.96\textwidth]{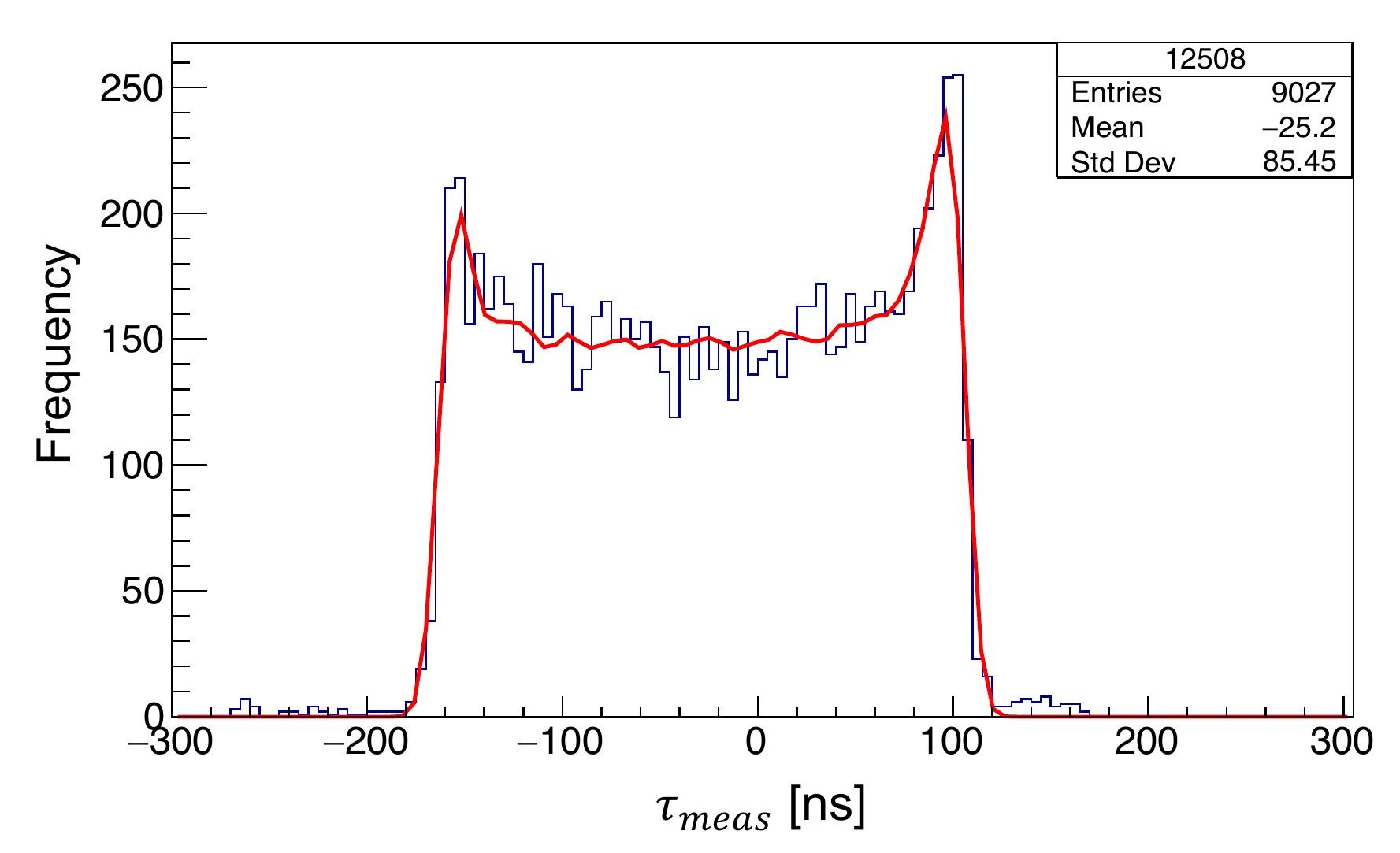}
            \caption{The CTD template yields a good match to the data for pixel 12508 in detector 1.}
            \label{fig:depth_CTDfit}
    \end{subfigure}
    \begin{subfigure}{0.49\textwidth}
            \includegraphics[width=\textwidth]{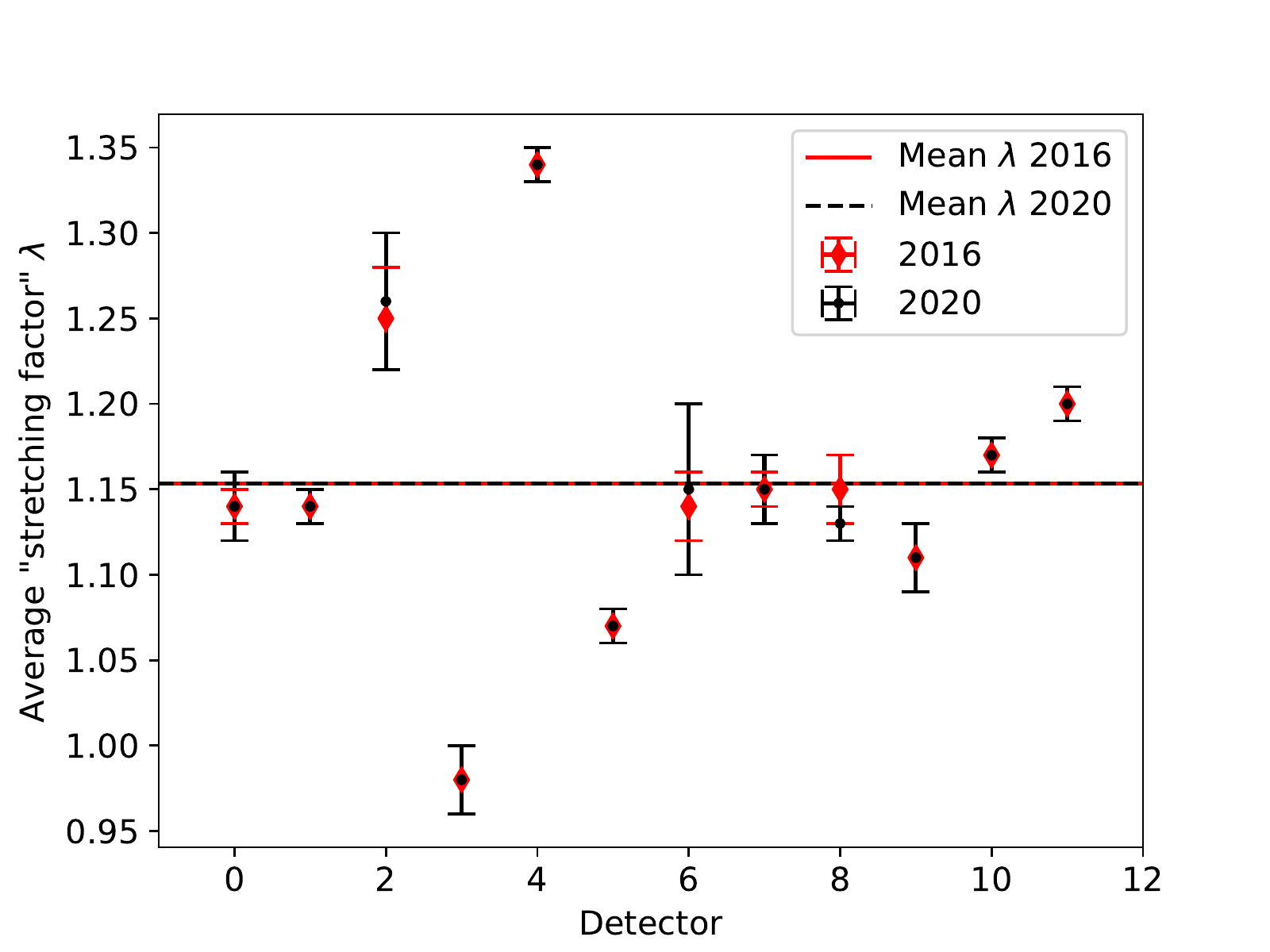}
            \caption{Average $\lambda$}
            \label{fig:depth_average_lambda}
    \end{subfigure}

    \begin{subfigure}{0.49\textwidth}
            \includegraphics[width=\textwidth]{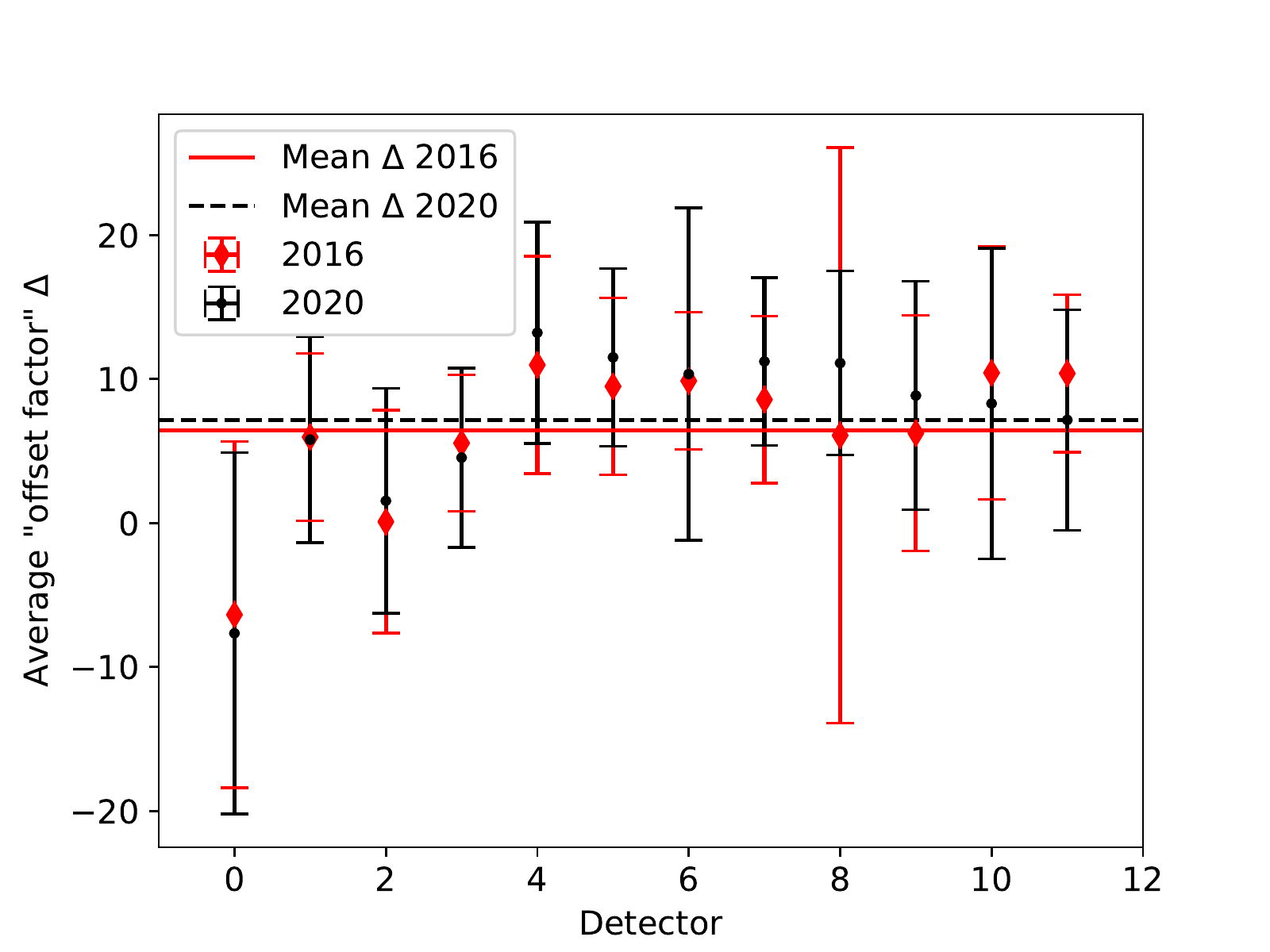}
            \caption{Average $\Delta$}
            \label{fig:depth_average_delta}
    \end{subfigure}
    \begin{subfigure}{0.49\textwidth}
            \includegraphics[width=\textwidth]{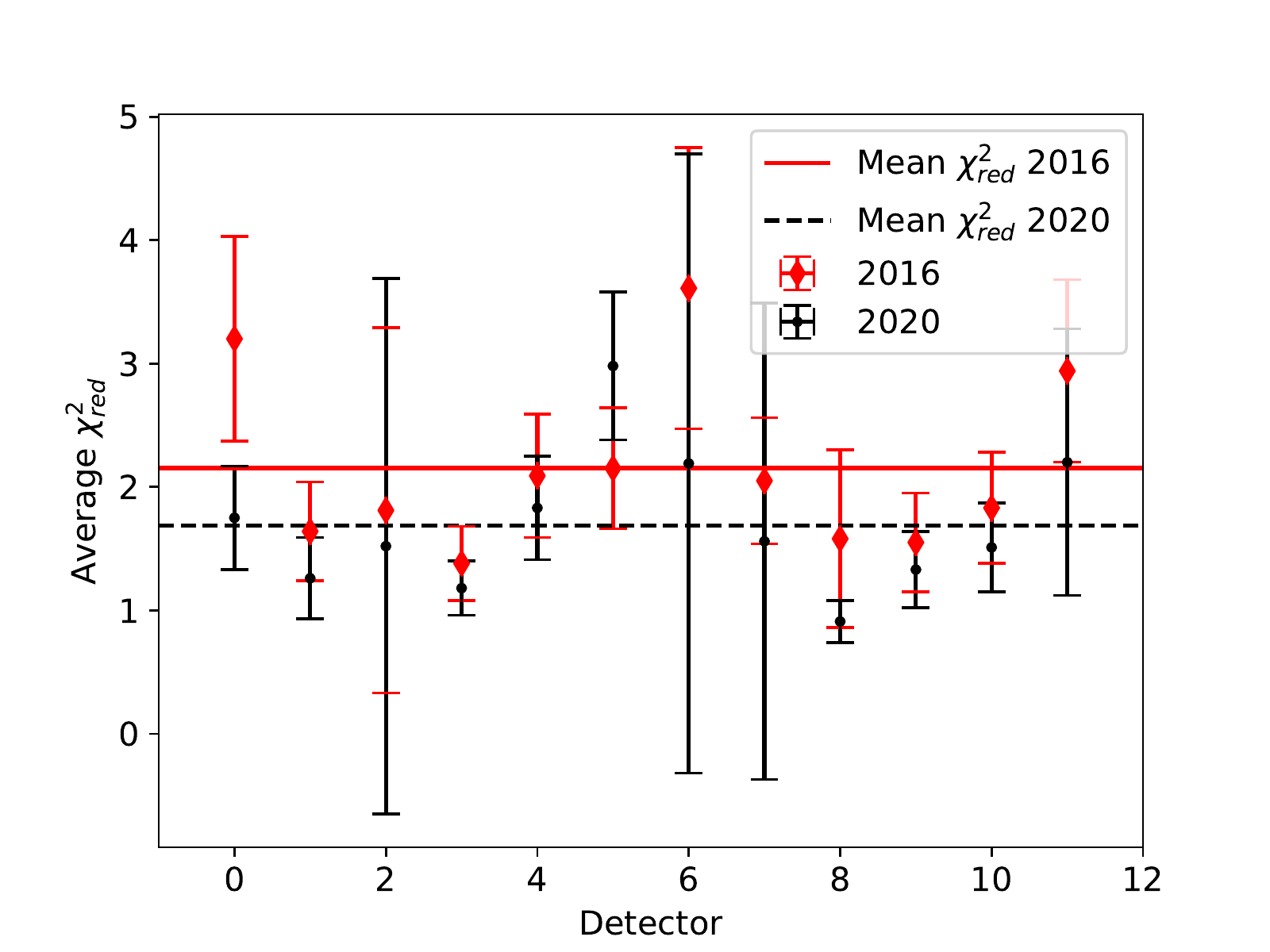}
            \caption{Average reduced $\chi^2$}
            \label{fig:depth_average_chisq}
    \end{subfigure}
    \caption{Depth calibration from 2016 and 2020. (a) 2020 example of a measured CTD (blue) and its corresponding fit (red) as generated with the CTD template. The CTD template yields a good match to the data. Also shown are comparisons between the 2016 and 2020 depth calibrations' (b) mean stretching factor, (c) offset factor, and (d) reduced $\chi^2$ for each detector, where the average is taken over all calibrated pixels in the detector of interest. Symmetric error bars indicate one standard deviation spread in value as averaged across all pixels in each detector.}
    \label{fig:depth}
\end{figure*}

As a measure of consistency in the calibration approach and performance of the detectors between 2016 and 2020, we compare each year's transformation factors $\lambda$ and $\Delta$ and the reduced $\chi^2$ of the CTD templates to measured CTDs, averaged over the pixels in each detector (Figure\,\ref{fig:depth}). 

The stretching and offset factors in 2016 and 2020 are largely consistent with each other. The reduced $\chi^2$ values differ more noticeably. The mean 2016 reduced $\chi^2$, averaged across all pixels and detectors, is 2.2 and the mean 2020 reduced $\chi^2$ is 1.7. 
The precise reason for the apparent improvement in CTD template fits to the data is uncertain, but improvements made to the DEE \cite{SLEATOR2019162643} after 2016 likely aided the success of the calibration. Additionally, we have made improvements to Cosima and MEGAlib in their development which may have enhanced our ability to mimic experimental conditions more accurately. In both 2016 and 2020, the detectors at the top of the COSI cryostat (0, 5, 6, 11) have greater mean reduced $\chi^2$ values than the other detectors. Detectors at the bottom of the cryostat (2, 3, 8, 9) have among the lowest reduced $\chi^2$ values. It is suspected that the higher count rate at the top detectors suppresses contribution of statistical noise in favor of higher systematic uncertainties.

\subsubsection{Future improvements}
Improvements to the depth calibration are being developed for future applications. For example, a new approach to the depth calibration, explained in detail in \cite{lowell:2016} and \cite{lowell:phdthesis}, promises more robust treatment of inhomogeneities in the detectors. Rather than using the same $z_n({\tau_{\rm sim}})$ for all pixels in each detector $n$, the new approach treats each pixel $p$ individually with unique lookup tables $z_p({\tau_{\rm meas}})$. To better understand these inhomogeneities, we plan to combine more comprehensive charge transport simulations (modeling all $37 \times 37$ strips rather than the $5 \times 5$ simplification used above) with recent advancements in the DEE to capture small-scale physical detector effects. Finally, we are exploring the energy dependence of the depth calibration and by extension, how the timing resolution of COSI varies with interaction energy.

\subsection{Detector Effects Engine}
\label{sec:dee}

Simulations are required to understand COSI's expected response to calibration and flight data. Comparing results from simulated data to real measurements is referred to as benchmarking and informs predictions of instrument performance. In order to generate simulations which reflect real data, we must apply imperfections intrinsic to measurements in the COSI detectors to simulated data. This conversion of simulated events to simulated events which mimic real data is accomplished with the detector effects engine (DEE). 

The simulations generate hits with 3-D positions and energies (in physical units) as would be observed in an ideal detector. To match these simulations more closely with real data, the DEE inverts the calibrations discussed previously to store the simulated hits in terms of strip hit, detector ID, timing, and ADC. The DEE artificially applies real-life phenomena such as charge sharing, charge loss, and cross-talk in the detectors to the simulations. It also vetoes GeD events coincident with shield events, discards events on dead strips in the instrument, and ignores events that occur within a defined dead time of the electronics. The DEE thus transforms simulation data into data which resemble those collected under the influence of imperfections intrinsic to the detectors and readout electronics.

After treatment by the DEE, the simulation data are run through the event calibration pipeline (Figure\,\ref{fig:pipeline}) identically to the real data. Comparing simulation data processed by Nuclearizer to real measurements helps identify inaccuracies in our pipeline. Properly benchmarking the DEE in this way also allows for an accurate determination of instrument performance using simulated data, enabling predictions of COSI's response to astrophysical sources with greater confidence. A complete description of the DEE and demonstration through these extensive benchmarking tests that the DEE successfully models real detector effects are provided in \cite{SLEATOR2019162643}.

\section{Instrument Performance}
\label{sec:instrument_performance}

Prior to the 2016 balloon launch, the COSI team collected more than 200 separate calibration measurements over approximately 1500 hours. The calibration data taken prior to the attempted 2020 launch, though limited to 10 days, serve as a valuable check of instrument performance over years of operation. Understanding instrument performance from these calibrations is an iterative process and both 2016 and 2020 data are subject to continuous study as analysis techniques are developed with time. As such, the benchmarking results presented below are a cumulative reflection of COSI's most recently determined capabilities. As there were no design changes, the instrument's performance between the campaigns was expected and proved to be consistent. 

\subsection{Energy Resolution}
\label{sec:energy_resolution}

We examine the energy resolution of fully reconstructed Compton events to understand the instrument's spectral performance. Note that this energy resolution differs from the single-strip energy resolution discussed in Section\,\ref{sec:energy_calib}. The single-strip resolution considers hits on individual strips rather than the total energy resolution of a reconstructed event with several energy depositions across multiple detectors. Thus, the single-strip resolution is more a measure of GeD spectral performance while the resolution of fully reconstructed events informs the spectral performance of the instrument as a whole. Additionally, studying the fully reconstructed energy resolution can help to assess the fidelity of the combined calibration procedures discussed previously. 

Event reconstruction of multi-site events is performed in the MEGAlib pipeline (Figure\,\ref{fig:pipeline}): The energy calibration from Section\,\ref{sec:energy_calib} converts ADC to energy, the temperature and cross-talk corrections are applied, the strip-pairing algorithm determines the most likely $x$--$y$ positions of the interactions, and the depth calibration returns the $z$-coordinate of the interactions. Using these fully reconstructed Compton events from 2020 calibration data, we calculate the energy resolution as the ratio of a reconstructed photopeak's FWHM to its true line energy. 

Prior to cancellation of the 2020 campaign, calibration data from $\mathrm{^{137}Cs}$, $\mathrm{^{60}Co}$, and $\mathrm{^{22}Na}$ were collected for this purpose. The resolutions in 2020 calibration data are comparable to those in 2016 data (Table\,\ref{table:reconst_energy_res}). The latter are as reported in \cite{SLEATOR2019162643}. 

\begin{table}[htbp]
\centering
\caption{Fully-reconstructed energy resolution in 2020 and 2016 \cite{SLEATOR2019162643} calibration data.}
\resizebox{\textwidth}{!}{%
\begin{tabular}{cccccc}
Isotope & \begin{tabular}[c]{@{}c@{}}Line energy \\ {[}keV{]}\end{tabular} & \begin{tabular}[c]{@{}c@{}}2020 FWHM \\ {[}keV{]}\end{tabular} & \begin{tabular}[c]{@{}c@{}}2020 Reconstructed \\ energy resolution {[}\%{]}\end{tabular} & \begin{tabular}[c]{@{}c@{}}2016 FWHM \\ {[}keV{]}\end{tabular} & \begin{tabular}[c]{@{}c@{}}2016 Reconstructed \\ energy resolution {[}\%{]}\end{tabular} \\
\hline
\hline
$\mathrm{^{22}Na}$ & 511.0 & $5.78 \pm 0.01$ & 1.1 & $5.56 \pm 0.04$ & 1.1 \\
$\mathrm{^{137}Cs}$ & 661.7 & $5.27 \pm 0.01$ & 0.8 & $5.1 \pm 0.02$ & 0.8 \\
$\mathrm{^{60}Co}$ & 1173.2 & $6.80 \pm 0.02$ & 0.6 & $7.36 \pm 0.05$ & 0.6 \\
$\mathrm{^{22}Na}$ & 1274.5 & $7.04 \pm 0.03$ & 0.6 & $6.42 \pm 0.1$ & 0.5 \\
$\mathrm{^{60}Co}$ & 1332.5 & $6.97 \pm 0.02$ & 0.5 & $6.95 \pm 0.05$ & 0.5 \\
\hline
\end{tabular}
}
\label{table:reconst_energy_res}
\end{table}

\subsection{Angular Resolution}
\label{sec:angular_resolution}

\begin{figure}\centering
\subfloat[]{\label{fig:arm_circles}\includegraphics[width=.49\linewidth]{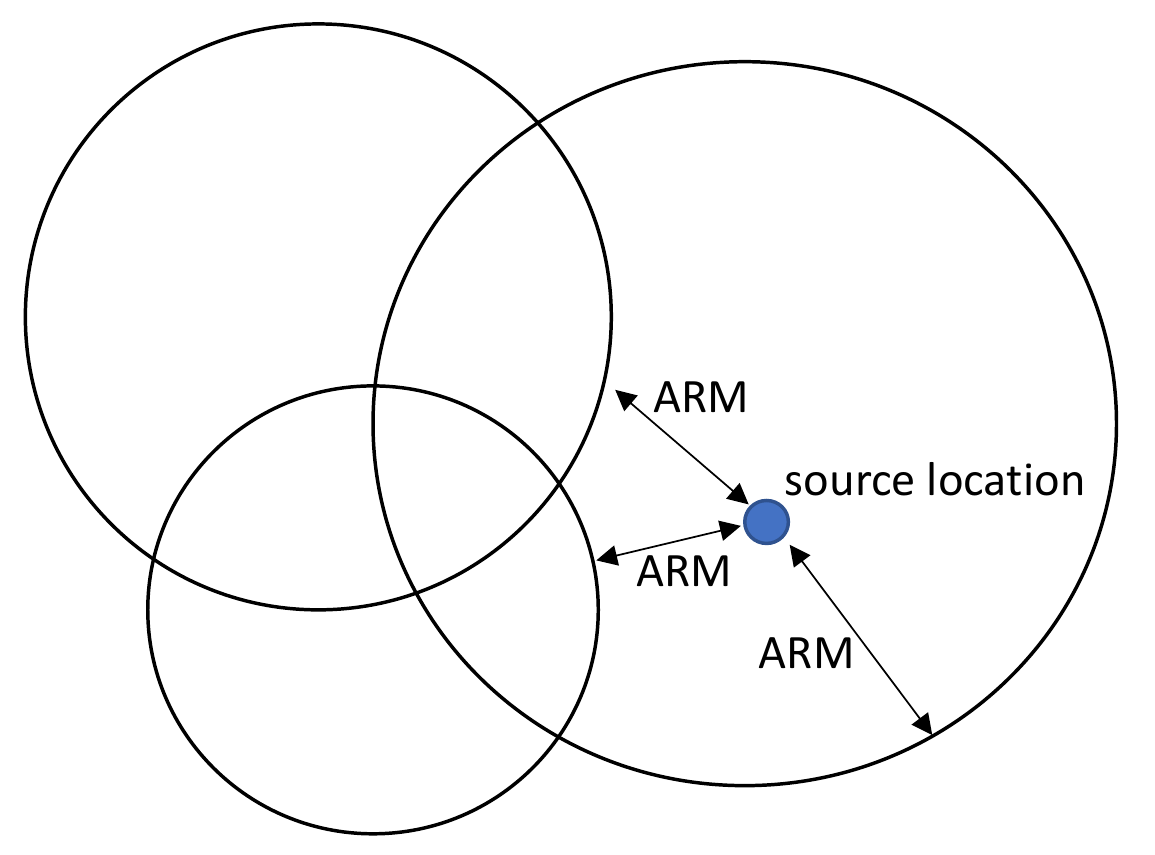}}\hfill
\subfloat[]{\label{fig:Cs137_ARM_2020}\includegraphics[width=.49\linewidth]{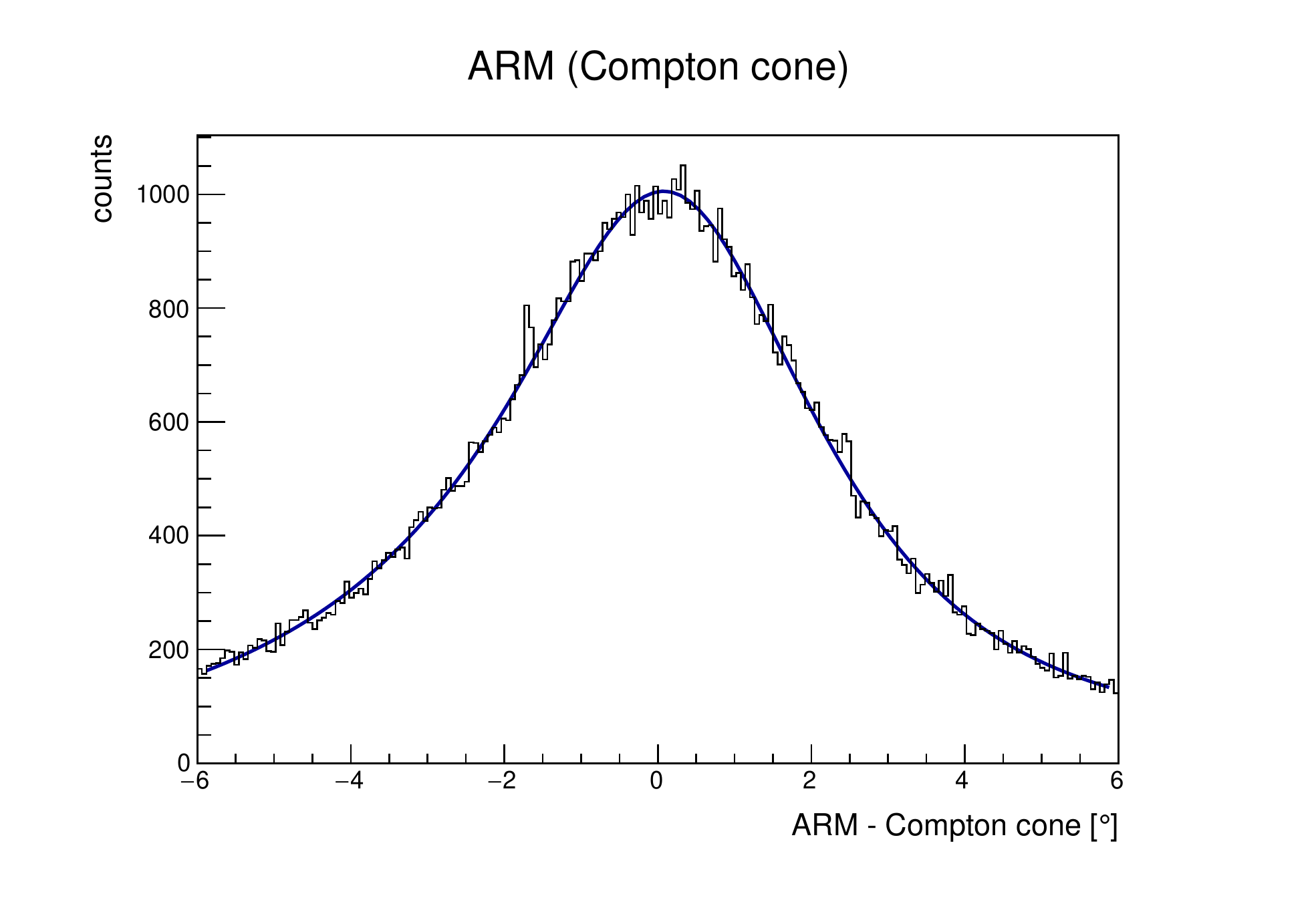}}\par
\subfloat[]{\label{fig:angular_resolution}\includegraphics[width=.49\linewidth]{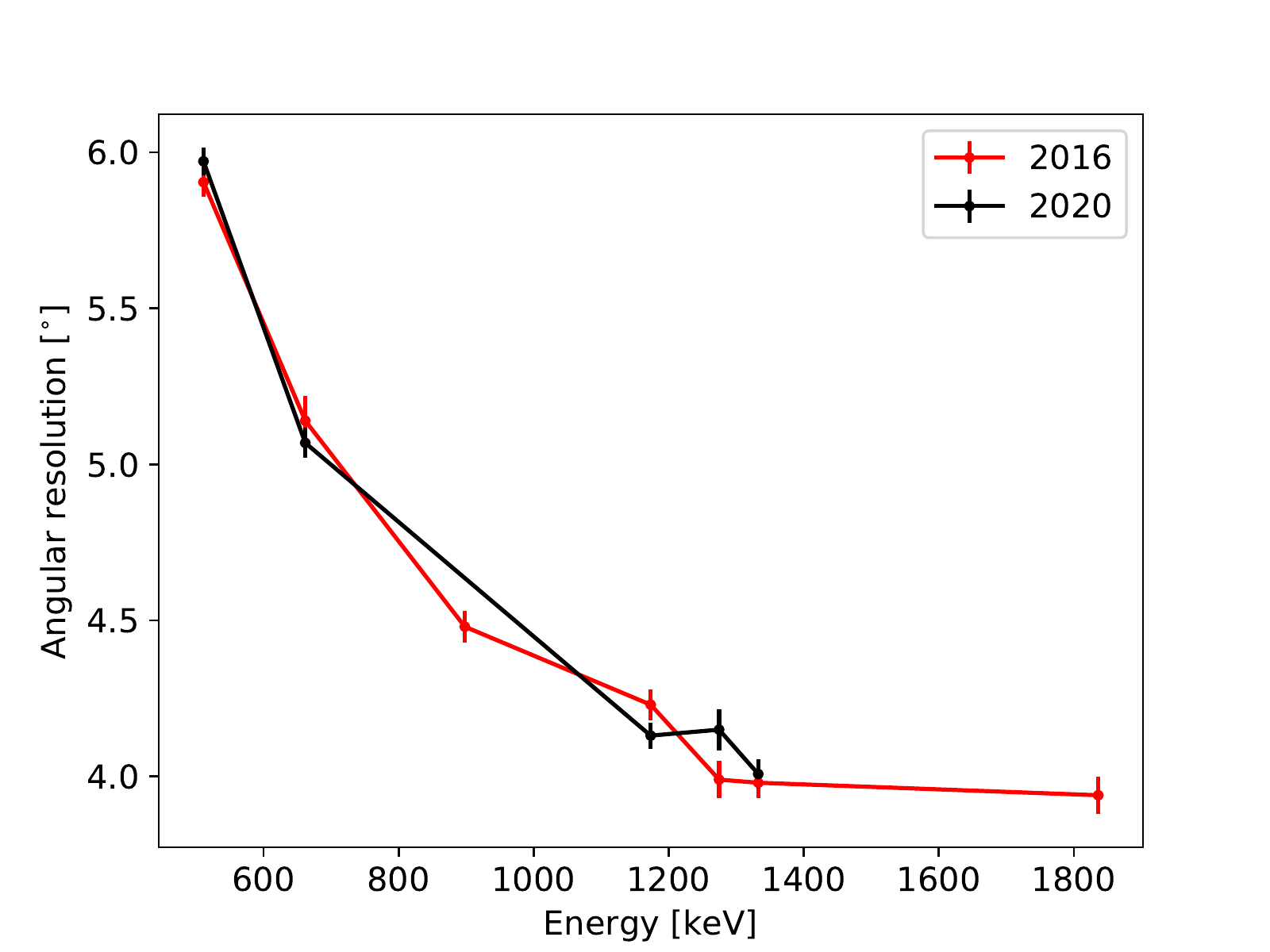}}
\caption{(a) An illustration of the angular resolution measure (ARM). The ARM of each event is the smallest angular distance between the known source location (blue dot) and the event circle (black circle). (b) The distribution of ARM values for 2020 $\mathrm{^{137}Cs}$ data. The FWHM of the double Lorentzian plus asymmetric Gaussian fit to the central peak of the ARM distribution defines the angular resolution. (c) The angular resolution as a function of energy in 2016 and 2020 calibration data.}
\label{fig:ARM}
\end{figure}

The angular resolution of a Compton telescope is defined by the FWHM of the angular resolution measure (ARM) distribution. For a given sample of Compton events, the ARM is the smallest angular distance between a source's known location and the event circle of each event (Figure\,\ref{fig:arm_circles}). The full distribution of ARM values represents the effective width of the telescope's point spread function, and its FWHM is an estimate of the angular resolution. In theory, the angular resolution is fundamentally limited by the Doppler broadening of the scattering electron \cite{du1929compton}, which is neither free nor at rest as assumed in the Compton equation. In practice, the angular resolution is limited by the accuracy of position and energy measurements. The dominant uncertainty is the 3-D position of each interaction, which is in turn dominated by the strip pitch of the detectors.

We calculate the angular resolution of COSI using the 2020 calibration data by measuring the ARM FWHM of $\mathrm{^{60}Co}$, $\mathrm{^{137}Cs}$, and $\mathrm{^{22}Na}$ line emissions. Angular resolution measurements at 898\,keV and 1836\,keV from $\mathrm{^{88}Y}$ data were taken in 2016 and are included as additional points of reference. In 2020 we were limited only to measurements taken with sources directly overhead the cryostat, at zenith. Using the Gaussian width $\sigma$ measured in Section\,\ref{sec:energy_resolution}, we select only the photopeak events by applying an energy cut of $\pm$1.5$\sigma$ to each line emission. Additional cuts are applied to restrict the Compton scattering angle to less than 90$^{\circ}$, impose a minimum distance between any two interactions of 1\,cm, and reject events originating 90$^{\circ}$ from COSI's zenith. The ARM FWHM is highly dependent on these event selections; the above are empirically designed to optimize the FWHM. The central peak of the ARM distribution (e.g. $\pm$6$^{\circ}$ at 662 keV)
was fitted with a double Lorentzian plus asymmetric Gaussian function to determine the FWHM. The distribution and fit to 2020 $\mathrm{^{137}Cs}$ data is shown in Figure \ref{fig:Cs137_ARM_2020} as an example.

The angular resolution as a function of $\gamma$-ray energy for the 2016 and 2020 calibration measurements is shown in Figure\,\ref{fig:angular_resolution} and Table\,\ref{table:angular_resolution}. The results from 2016 are a re-analysis of 2016 data; a similar analysis was originally published in \cite{SLEATOR2019162643}. The resolutions in 2016 and 2020 data are largely consistent over the tested energy range. We expect improved angular resolution (smooth, monotonically decreasing ARM values) with increasing energy because higher incident photon energy increases the distance between Compton interactions, which improves the accuracy of event reconstruction. This trend is visible in both sets of data. The angular resolution behaves as expected with energy and is consistent between 2016 and 2020 measurements.

\begin{table}[htbp] 
\centering
\caption{Angular resolution of 2020 calibration data compared to that of 2016. Event selections: Compton events with incident photon energy within $\pm$1.5$\sigma$ of the photopeak line energy, Compton scattering angle 0$^{\circ}$ to 90$^{\circ}$, 2--7 total interactions, minimum distance between any two interactions of 1\,cm.}
\resizebox{\textwidth}{!}{%
\begin{tabular}{cccccc}
Isotope & Line energy [keV] & 
2020 Angular resolution [$^{\circ}$] & 2016 Angular resolution [$^{\circ}$] \\ 
\hline
\hline
$\mathrm{^{22}Na}$ & 511.0 & $5.97 \pm 0.04$ & $5.9 \pm 0.1$ \\
$\mathrm{^{137}Cs}$ & 661.7 & $5.1 \pm 0.1$ & $5.1 \pm 0.1$ \\
$\mathrm{^{88}Y}$ & 898.0 & -- & $4.5 \pm 0.1$ \\
$\mathrm{^{60}Co}$ & 1173.2 & $4.13 \pm 0.04$ & $4.2 \pm 0.1$ \\
$\mathrm{^{22}Na}$ & 1274.5 & $6.1 \pm 0.3$ & $6.5 \pm 0.3$ \\
$\mathrm{^{60}Co}$ & 1332.5 & $4.2 \pm 0.1$ & $4.0 \pm 0.1$ \\
$\mathrm{^{88}Y}$ & 1836.0 & -- & $3.9 \pm 0.1$ \\
\hline
\end{tabular}
}
\label{table:angular_resolution}
\end{table}

\subsection{Effective Area}
\label{sec:effective_area}

The effective area of a telescope is a measure of its geometrical size and detection efficiency. It is the area that a photon sees when crossing the path of the instrument, weighted by effects including interaction probabilities, detector geometry, and event selections applied to the data. Calculating the effective area of a telescope over its energy bandpass and field of view is a crucial measure of instrument performance.

Effective area is defined as the product of the telescope's collecting area and its efficiency $\epsilon$ at a specified energy, where efficiency measures the fraction of all incident photons detected by the instrument. The measured photon rate, $L_{\rm meas}$, is the number of photons in the measured photopeak, $N$, divided by exposure time $t_{\rm meas}$. The incident photon rate, $L_{\rm inc}$, is the product of the source activity $S$, the branching ratio BR of photon emission at the given energy, and the detector area divided by $4\pi z^2$, for a source at distance $z$ from the detector. The latter term represents the surface area subtended by our instrument of the $4\pi z^2$ sphere surrounding the source. Air attenuation is defined by the linear attenuation coefficient, $\lambda$, of air at the given energy. Additionally, we correct for the dead time of the instrument during data collection. The analog boards and coincidence logic take $\sim$10\,$\mu$s to execute after each triggered event. Any hits which occur during this 10\,$\mu$s dead time are not recorded. The calibration sources are of sufficient activity to induce non-negligible dead time, which we correct for by scaling the number of photons in the measured photopeak by a factor $\delta$.

Multiplying the efficiency, $\epsilon$, by the geometric area of the detector, $A_{\rm geo}$, gives the effective area $A_{\rm eff}$:

\begin{gather*}
    A_{\rm eff} = A_{\rm geo} \cdot \epsilon \\
    \epsilon = L_{\rm meas} / L_{\rm inc} \\
    L_{\rm meas} = \frac{N \cdot \delta}{t_{\rm meas}}, \text{for } \delta = \frac{100}{100 - \textrm{dead time [\%]}} \\
    L_{\rm inc} = S \cdot \textrm{BR} \cdot \frac{A_{\rm geo}}{4\pi z^2} \cdot e^{-\lambda z} 
\end{gather*}

Then,

\begin{equation}
    A_{\rm eff} = 4\pi z^2 \frac{N \cdot \delta}{t_{\rm meas} \cdot S \cdot \textrm{BR} \cdot e^{-\lambda z}}
    \label{eq:effective_area}
\end{equation}

\begin{figure}[htbp]
\centering
\includegraphics[width = 0.7\textwidth]{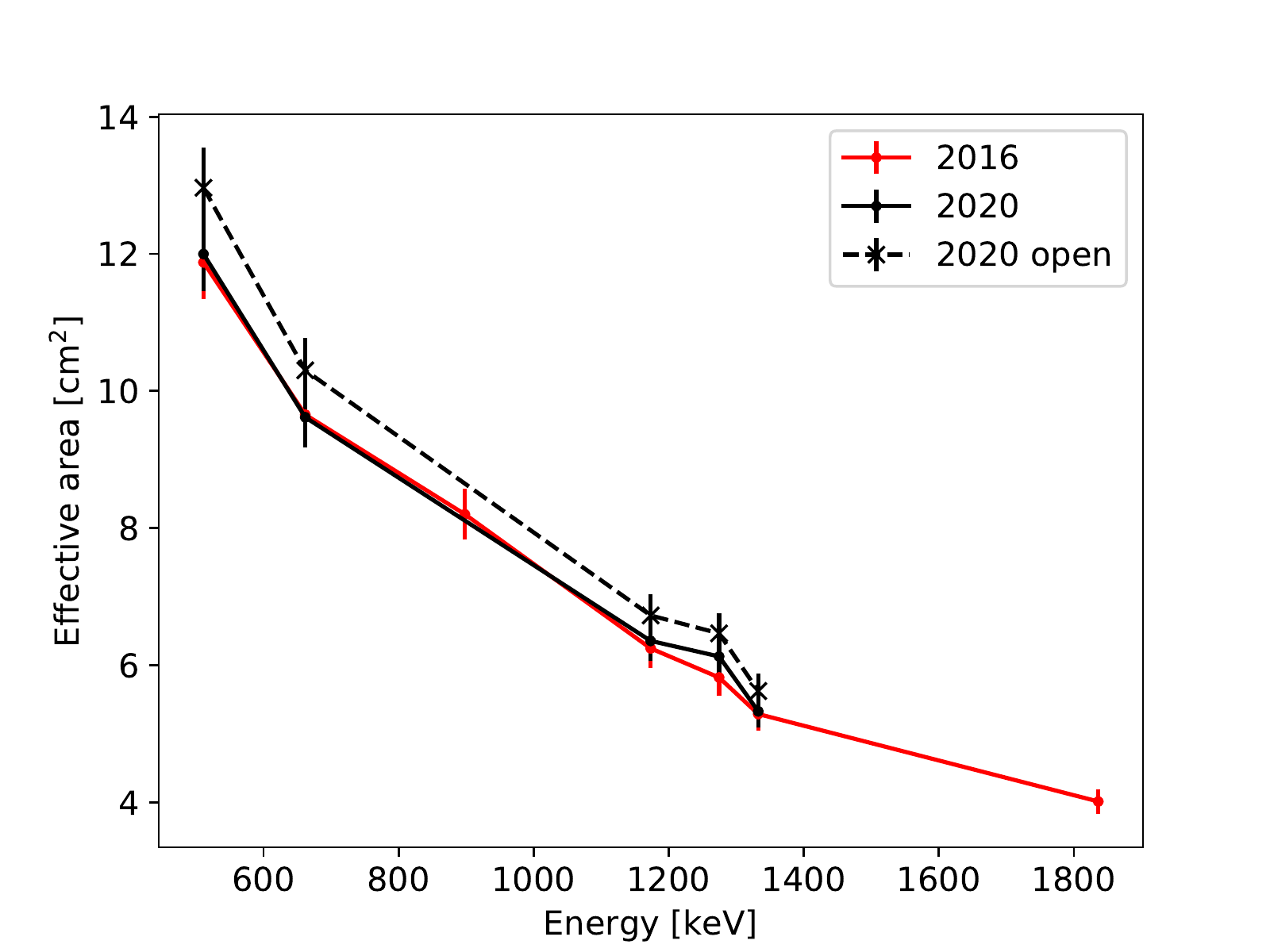}
\caption{The effective area as a function of energy in 2020 and 2016 calibration data. The error bars include statistical uncertainties and systematic uncertainties in source activity and distance from the detector.}
\label{fig:eff_area}
\end{figure}

We calculate the effective area of COSI using 2020 and 2016 calibration measurements. Attenuation in air at these energies and distances is negligible. We consider Compton events which pass the following event selections: incident photon energy within $\pm$2$\sigma$ of the photopeak line energy, Compton scattering angle 0--180$^{\circ}$, 2--7 total interactions, minimum distance between the first two interactions of 0.5\,cm, and minimum distance between any two interactions of 0.3\,cm. Figure\,\ref{fig:eff_area} and Table\,\ref{table:effective_area} show the effective area as a function of incident photon energy in 2020 and 2016 calibration data. The values are consistent between years.

In the 2020 data, we relax the minimum distances between interactions to 0\,cm to demonstrate the dependence of the effective area on event selections. This leads to a $\sim$8.1\% greater effective area at 511\,keV and $\sim$6.4\% greater effective area averaged across the photopeaks, with decreasing improvement as the photon energy increases. Note that this example of decreasing the minimum distance between interactions worsens angular resolution; curating event selections to the analysis task at hand often requires careful consideration of competing performance metrics.

In general, we observe the expected trend of decreasing effective area with increasing energy. As incident photon energy increases, so does the probability that a scattered photon escapes the detector volume. Such incompletely absorbed events cannot be reconstructed, thereby limiting COSI's effective area.

\begin{table}[htbp]
\centering
\caption{Effective area of 2020 calibration data compared to that of 2016 (Figure\,\ref{fig:eff_area}). Event selections: Compton events with incident photon energy within $\pm$2$\sigma$ of the photopeak line energy, Compton scattering angle 0--180$^{\circ}$, 2--7 total interactions, minimum distance between the first two interactions of 0.5\,cm, and minimum distance between any two interactions of 0.3\,cm. ``Open" event selections on 2020 relax the minimum interaction distances to 0\,cm. The error bars include statistical uncertainties and systematic uncertainties in source activity and distance from the detector.}
\resizebox{\textwidth}{!}{%
\begin{tabular}{cccccc}
Isotope & Line energy & 
2020 Effective area & 2020 Effective area (``open") & 2016 Effective area \\ 
& [keV] & [cm$^{2}$] & [cm$^{2}$] & [cm$^{2}$] \\
\hline
\hline
$\mathrm{^{22}Na}$ & 511.0 & $12.0 \pm 0.5$ & $13.0 \pm 0.6$ & $11.9 \pm 0.5$ \\
$\mathrm{^{137}Cs}$ & 661.7 & $9.6 \pm 0.4$ & $10.3 \pm 0.5$ & $9.7 \pm 0.4$ \\
$\mathrm{^{88}Y}$ & 898.0 & -- & -- & $8.2 \pm 0.4$ \\
$\mathrm{^{60}Co}$ & 1173.2 & $6.4 \pm 0.3$ & $6.7 \pm 0.3$ & $6.3 \pm 0.3$ \\
$\mathrm{^{22}Na}$ & 1274.5 & $6.1 \pm 0.3$ & $6.5 \pm 0.3$ & $5.8 \pm 0.3$ \\
$\mathrm{^{60}Co}$ & 1332.5 & $5.3 \pm 0.2$ & $5.6 \pm 0.3$ & $5.3 \pm 0.2$ \\
$\mathrm{^{88}Y}$ & 1836.0 & -- & -- & $4.0 \pm 0.2$ \\
\hline
\end{tabular}
}
\label{table:effective_area}
\end{table}

\subsection{Polarization Response}
\label{sec:polarization}
 
In order to determine COSI’s polarization response and to identify systematic deviations from an ideal sinusoidal modulation, it is necessary to evaluate COSI's polarization performance in the laboratory. In 2019 at SSL, we collected data from partially-polarized beams and ran simulations that mimic the experimental configuration. We also produced simulations of unpolarized sources in order to capture geometric effects of the instrument.

Compton telescopes preserve information from linearly polarized photons in a relationship described by the Klein-Nishina equation. The polarization fraction and angle can be derived from

\begin{equation}
    P(\nu) = A \cdot \cos (2 (\nu- \nu_{0})) + P_{0},
\end{equation}

\noindent where A is the amplitude, $P_{0}$ is the offset, $\nu$ is the azimuthal scattering angle, and $\nu_{0}$ is the polarization angle. The measured azimuthal scattering angle distributions (ASADs), which have a characteristic sinusoidal shape, allow us to infer the polarization angle and level from a polarized source. To correct for geometric and systematic effects, we generate ASADs for each run by simulating an unpolarized source in the same location of the scintillator.  

To produce an ASAD of the partially-polarized beam of coincident events, we use a series of event selections determined in \cite{lowell:phdthesis}. These include:

\begin{itemize}
  \item $\pm 1 \sigma$ of the photopeak energy, obtained by performing a simple Gaussian fit to the photopeak.
  \item $\pm 1 \sigma$ of the FWHM of the ARM distribution, which was generated for the photopeak events. 
  \item Distance between first two interactions of 1.0\,cm, and a minimum distance between any two interactions of 0.5\,cm.
\end{itemize}

\begin{figure}[htbp]
\centering
\includegraphics[width = 0.9\textwidth]{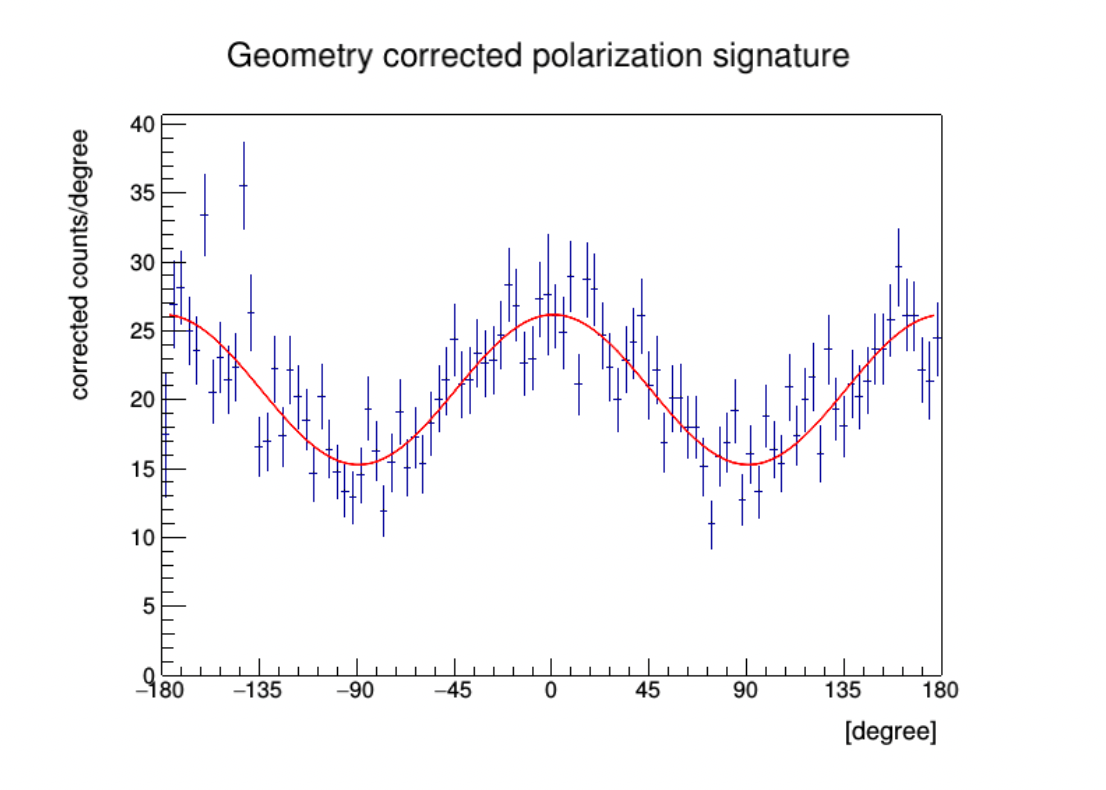}
\caption{The geometrically-corrected ASAD with best fit modulation curve using the listed event selections.}
\label{fig:asad_fin}
\end{figure}

\noindent The first two selections reduce background and the last selection omits events with interactions close to one another, thereby reducing effects such as charge sharing, charge loss, and cross-talk. We consider the ASAD after subtracting the chance coincident background and geometrically correct it using the unpolarized ASAD obtained from a simulation. By subsequently applying the listed event selections, we obtain the final ASAD pictured in Figure\,\ref{fig:asad_fin}. This ASAD provides an important validation that the corrected ASAD is sinusoidal and has the correct phase for the known polarization angle. 

The measured calibration data are then compared to simulations to ensure a good match between the two and if not, adapt the simulations and the DEE to the data until a good match is found. This process is called benchmarking. The Kolmogorov-Smirnov (KS) test was used to evaluate the hypothesis that the azimuthal scattering angle samples from the measurement and simulations were drawn from the same underlying distribution at a 99\% confidence level. 

\begin{figure}[htbp]
\centering
\includegraphics[width = 0.8\textwidth]{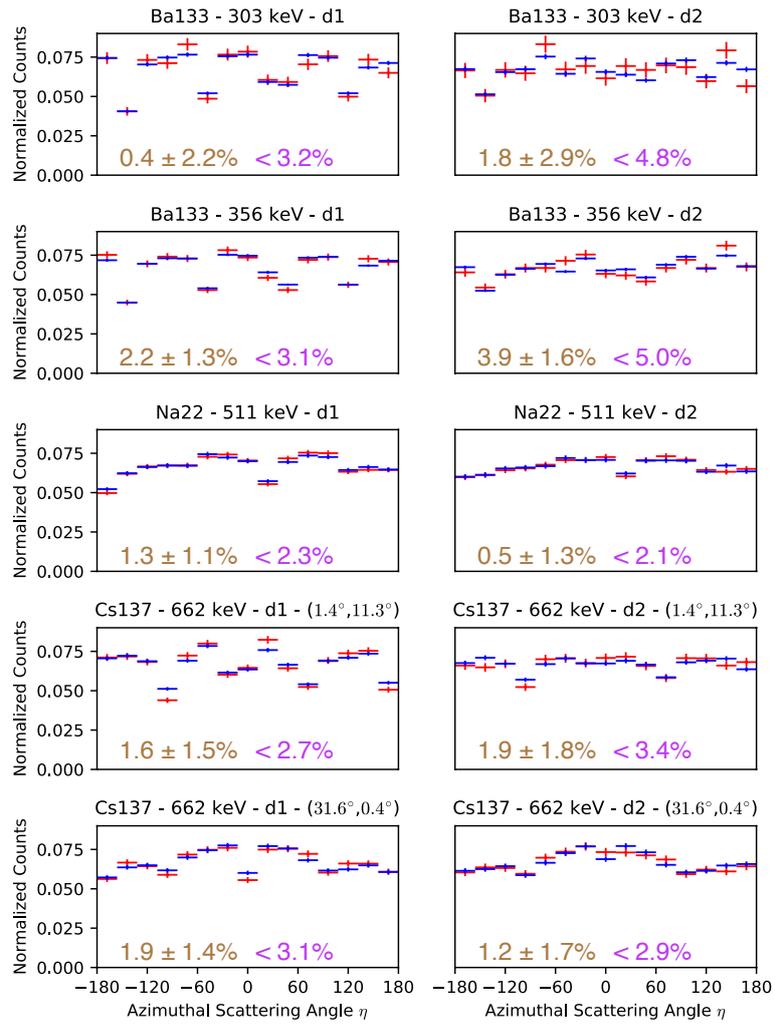}
\caption{The modulation amplitude measured for unpolarized sources of different energies. Adapted from \cite{lowell2017}.}
\label{fig:sys_unc}
\end{figure}

The residual systematics between simulations and calibrations were characterized at the $\sim$2--5\% level throughout the COSI field of view, as detailed in \cite{lowell:phdthesis}. Unpolarized data from the 2016 campaign were collected with radioactive sources placed throughout the instrument field of view, enabling characterization of the systematic modulations that ultimately could limit polarization sensitivity. First, KS tests and Anderson-Darling tests were applied to validate that the modulations of measured calibration data (blue in Figure\,\ref{fig:sys_unc}) and simulations (red in Figure\,\ref{fig:sys_unc}) are drawn from the same distributions. The modulations were then normalized to unity and fit to determine the amplitude and its uncertainty (brown in Figure\,\ref{fig:sys_unc}). Lastly, a bootstrapping analysis of $10^{4}$ samples was conducted to place an upper limit on the residual modulation (purple in Figure\,\ref{fig:sys_unc}).

This process has shown that our simulations are well-benchmarked with calibrations, and therefore these benchmarked simulations can be used to create full all-sky response files for, e.g., imaging and polarization analysis.

\section{Conclusion}
\label{sec:conclusion}

In this paper we detailed the calibration procedures of COSI and found consistent instrument performance between 2016 and 2020. The energy resolution of COSI's strips is $\sim$0.5\%, the angular resolution $\sim$5.1$^{\circ}$, and the effective area $\sim$9.6\,cm$^2$ at 661.7\,keV. The notable science portfolio from the 2016 flight, spanning the 511\,keV Galactic annihilation line, GRB polarization, measurement of Galactic $\mathrm{^{26}Al}$, and imaging of compact objects, is a validating manifestation of these procedures. COSI's demonstrated capabilities on a balloon platform firmly establish the potent ability of Compton telescopes to probe the MeV gap of $\gamma$-ray astrophysics. The next generation of COSI as a NASA Small Explorer satellite mission \cite{tomsick2019compton} is scheduled for launch in 2025 and will require similar calibration and benchmarking to realize its rich scientific potential.

\section{Acknowledgements}
We thank engineers Brent Mochizuki and Steve McBride for their contributions to COSI's electronics and data acquisition system. 
\\
\\
\noindent Software: matplotlib \cite{Hunter:2007}, MEGAlib \cite{zoglauer2006megalib}, numpy \cite{harris2020array}, scipy \cite{2020SciPy-NMeth}.
\\
\\
\noindent Funding: This work was supported by NASA Astrophysics Research and Analysis USA grant 80NSSC19K1389 and the Centre National d'\'Etudes Spatiales (CNES). Thomas Siegert is supported by the German Research Foundation DFG-Forschungsstipendium SI 2502/3-1.


\bibliographystyle{elsarticle-num-names} 

\end{document}